\pdfoutput=1
\documentclass[acmtog, nonacm]{acmart}

\acmJournal{TOG}

\usepackage{bm}

\usepackage{wrapfig}
\usepackage{graphicx}
\usepackage{colortbl}
\usepackage{multirow}
\usepackage{tabularx}
\usepackage{algorithm}
\usepackage{algpseudocode}
\definecolor{oursrow}{HTML}{E2F2DF}  %
\definecolor{algcomment}{gray}{0.55}  %
\algrenewcommand{\algorithmiccomment}[1]{\hfill{\color{algcomment}\(\triangleright\) #1}}

\AtBeginDocument{%
  \theoremstyle{acmdefinition}%
  \theoremstyle{acmplain}%
}

\begin{document}

\title{Piecewise CMC Segmentation for Fluidic Shaping}

\author{Abhinit Sati}
\email{abhinit@bu.edu}
\orcid{}
\affiliation{%
  \institution{Boston University}
  \city{Boston}
  \state{Massachusetts}
  \country{USA}
}

\author{Jacob Garrett}
\email{jpg2002@bu.edu}
\affiliation{%
 \institution{Boston University}
  \country{USA}
 }

\author{Valeri Frumkin}
\email{valfru@bu.edu}
\affiliation{%
  \institution{Boston University}
  \country{USA}
}

\author{Edward Chien}
\email{edchien@bu.edu}
\affiliation{%
  \institution{Boston University}
  \country{USA}
}

\author{Emily Whiting}
\email{whiting@bu.edu}
\affiliation{%
  \institution{Boston University}
  \country{USA}
}

\begin{abstract}

  We present a novel mesh segmentation approach for the fabrication of 3D geometries using fluidic shaping. The physics of fluidic shaping gives rise to Constant Mean Curvature (CMC) surfaces due to interfacial tension between the polymer fluid used to shape the 3D structure and an immersion fluid of equal density.
  Our mesh segmentation algorithm partitions the mesh into a piecewise CMC surface, such that every segment is a CMC approximation to the target within a user specified tolerance.
  We run a CMC evolution procedure in the loop during segmentation to find physically stable CMC surfaces that can be reliably fabricated.
  We extract a wireframe scaffold from the seams of this segmentation and inject polymer fluid into the frame for curing.  
  The structure cures as a single, continuous, isotropic volume, yielding nanometer scale roughness on the interior surface of each segment without the costly post-processing or smoothing that conventional state-of-the-art approaches require for comparable surface quality.

\end{abstract}


\begin{teaserfigure}
  \includegraphics[width=\textwidth]{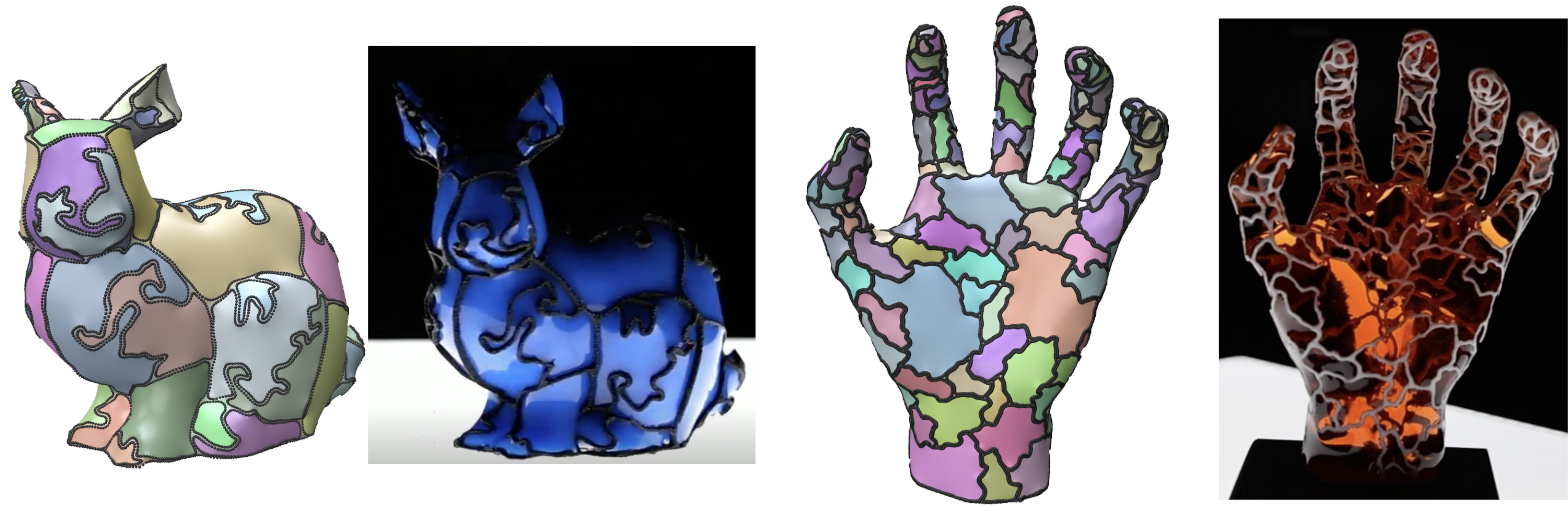}
  \caption{We use fluidic shaping to fabricate complex structures using the physics of interfacial phenomena. Our method takes as input a 3D shape, computes a piecewise Constant Mean Curvature segmentation and outputs a sparse bounding frame. The frame acts as a scaffold allowing polymer fluid to take the desired shape, which is eventually cured to get the final structure. }
  \label{fig:teaser}
\end{teaserfigure}

\maketitle

\begin{figure*}[!t]
  \centering
  \includegraphics[width=0.92\textwidth]{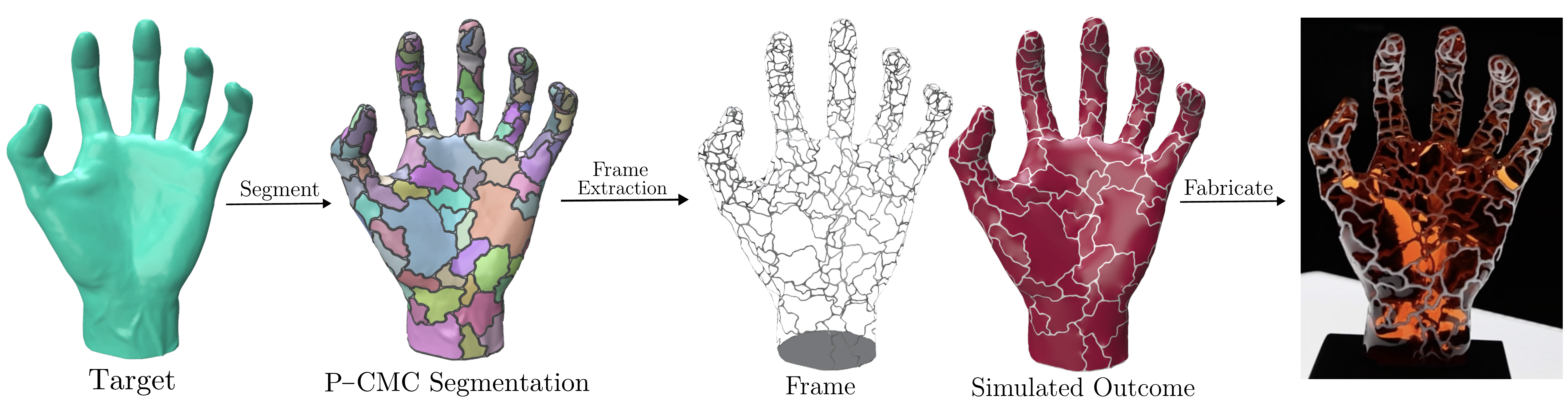}
  \caption{Our computational pipeline. We segment the input geometry into patches that can be approximated by CMC surfaces of a given target MC=$H^*$. The segment seams are used to extract a frame that is 3D printed. Finally, we use \textit{Fluidic Shaping} to manufacture the geometry, where the frame acts as a scaffold. }
  \label{fig:pipeline}
\end{figure*}

\section{Introduction}

Fluidic shaping \cite{frumkin2021fluidic, elgarisi2021fabrication} is a fabrication technique that exploits properties of surface tension to produce exceptionally smooth surfaces without the need for machining or polishing. Originally developed for optical components, the process involves constraining a liquid polymer to a rigid frame within an immiscible bath, which naturally relaxes to a smooth shape with nanometer-scale surface roughness after curing \cite{elgarisi2025fabrication}. However, because the resulting geometry is governed by the equilibrium of fluid interfaces, existing approaches have a limited design space of lens-like surfaces, with no general approach for fabricating freeform 3D shapes.

In this work, we introduce a novel mesh segmentation approach that enables extending fluidic shaping to freeform 3D surfaces. A key insight is that in a density-matched fluid bath, gravitational forces are suppressed, allowing interfacial tension to drive a liquid interface toward a Constant Mean Curvature (CMC) surface. By partitioning a target geometry into CMC patches and constraining the liquid along their boundaries, we can produce complex shapes via the fabrication approach of \citet{Garrett2026Fluidic} that would otherwise be inaccessible to conventional fluidic shaping. This leads to a new geometric decomposition problem, which we call piecewise CMC (P-CMC) segmentation.  Analogous to developable segmentation \cite{julius2005d, wang2008computing, zhao2022developability} and mold paneling \cite{eigensatz2010paneling, fu2010kset}, a P-CMC segmentation decomposes freeform geometry into simpler manufacturable primitives.

A central challenge is that the individual segments within a frame remain connected through a common body of polymer fluid. In a density-matched bath, the pressure difference across the two fluids is the same everywhere and hence all patches must share the same mean curvature $H^*$ at equilibrium. This shared curvature $H^*$ can be prescribed by controlling the injected polymer volume during fluidic shaping, but imposes a global constraint on the geometric decomposition.

\looseness=1
We formulate the computational problem to solve as follows---\textit{Given a target Mean Curvature (MC) $H^*$, compute a segmentation of the input geometry such that each segment admits an evolution into a CMC surface, with the approximation error between the evolved and original geometry bounded by a user-defined threshold.} We employ a greedy approach, iteratively \textit{growing} segments on the input surface. Each segment starts from a single mesh element and expands until the resulting patch can no longer be approximated by a CMC surface. The seams separating the segments are precisely where the surface fails to be CMC, so the segmentation concentrates the sharp features of the target geometry onto a sparse collection of curves. This is akin to the developability flow of \citet{stein2018developability} and more recently \citet{chen2026design} who optimize for flank milling manufacturability, where Gaussian curvature gets concentrated onto a sparse collection of curves on the surface.

Physically, those seam curves are what the fabrication process needs as boundary data. Interfacial tension acts to contract a liquid interface toward minimal area; left unconstrained, a fixed volume of polymer would relax to the shape that minimizes area for that volume---a sphere---retaining none of the target geometry. Pinning the polymer liquid along the seams allows the remaining bulk to settle into the CMC patches our segmentation prescribes (Fig.~\ref{fig:teaser}).
The resulting structure cures into a continuous volume with isotropic mechanical properties, compared to conventional additive manufacturing approaches where material is deposited and cured layer by layer, leaving micrometer scale deposition marks on the surface and an anisotropic bulk volume \cite{montalti2024enhancing} \cite{mason2026robust}.

We summarize the contributions made in the paper as follows:
\begingroup\setlength{\listisep}{2pt}
\begin{itemize}
  \item We introduce the first algorithm for piecewise CMC segmentation, decomposing a surface into patches that each evolve into a CMC surface of prescribed mean curvature within a user-defined error.

  \item A graph contraction approach for growing segments, aided by a CMC evolution procedure robust to arbitrary input geometry.

  \item An end-to-end pipeline from segmentation to 3D printable wireframe, taking fluidic shaping from single optical surfaces to freeform 3D geometry.
\end{itemize}
\endgroup

\section{Related Work}

\subsection{Evolution into CMC Surfaces -- Tools for Forward Solve}
\label{subsec:rw_cmc}

A fixed boundary and a prescribed mean curvature $H^*$ generally admit both small and large
solutions~\cite{struwe1988plateau}, such as the minor and major caps of a sphere,
and methods differ mainly in which of them they can reach.
\textit{Surface Evolver}~\cite{brakke1992surface} prescribes $H^*$ through a
pressure term and descends on the energy $E$ of Equation~\eqref{eq:energy_H}.
It converges only to its local minimizers, thus missing saddle points of $E$
such as the major cap, which are also stable at fixed fluid volume
(Appendix~\ref{app:stability}). Surface Evolver reaches such surfaces under a
volume constraint and can analyze their stability~\cite{brakke1996surface}, but
the curvature then becomes an output. The same holds for \citet{pan2012robust}
and \citet{yu2016constructing}, who weight the same discrete volume gradient by
a scalar ($t$ and $\lambda$ respectively), and for \citet{renka2015simple}, who
prescribes a target volume. Mean curvature is not monotone in any of these
parameters (Fig.~\ref{fig:tH}; \citet[Fig.~11]{yu2016constructing}), so it
cannot be set by tuning them. We use the energy of \citet{pan2012robust} as the
inner solver and recover control of $H^*$ by root-finding over its parameter.

\citet{dziuk2006finite} and \citet{chicco2016algorithm} take $H^*$ as a true
input and solve by Newton-type iterations, but the former is restricted to disc
topology with a boundary that slides along the prescribed curve, and the latter
to smooth isogeometric spline patches. Both, like Newton generally, converge to
nearby solutions regardless of physical stability (Fig.~\ref{fig:stability}).
\citet{brubaker2018continuation} traces entire families of stable and unstable
solutions by continuation, but must be initialized with a surface that is already
CMC over a rectangular parameter domain. \citet{polthier2002discrete} compute
the index of discrete CMC surfaces from the discrete Jacobi operator that serves
as our Newton Jacobian in Section \ref{subsec:cmc_evo}, and \citet{pinkall1993computing} treat the minimal case where
$H^* = 0$. Finally, \citet{pellis2021computational} optimize toward Weingarten
surfaces, a class containing CMC surfaces, near an input design with the
curvature relation as an output, and \citet{soliman2026constant} prescribe
tangent planes rather than a boundary curve. To the best of our knowledge, our procedure is the first
that takes an arbitrary triangle mesh with pinned boundary and a
scalar $H^*$ and upon convergence, returns a CMC surface of that curvature
that is stable at fixed volume, including at certain saddle points of $E$ (see Appendix~\ref{app:stability} for more details).

\subsection{Fabrication via Interfacial Phenomena}

Recent work on shaping complex structures using the interfacial properties of fluids has gained rapid traction. \citet{duraivel2023silicone} propose a technique that tunes the interfacial tension between the support material and printing fluid to "jam" it in place, which they use to fabricate thin complex 3D structures. In a similar vein, \citet{hajash2017large} use reusable supports in the form a liquid gel to 3D print large scale structures. \citet{jones2021bubble} harness interfacial flows in a curing elastomer — an injected air bubble whose drainage sets the film thickness to cast monolithic soft pneumatic actuators. \citet{forth2018reconfigurable} print water-in-oil structures held together by an interfacially jammed nanoparticle "supersoap," fixing liquid threads into stable, reconfigurable geometries.

\subsection{Fluidic shaping for lens making}

Similar to the work in the previous section, fluidic shaping is a fabrication approach that uses interfacial tension to shape fluids into 3D structures. A characteristic feature of this approach is that the interfacial tension between two fluids yields nanometer-scale surface smoothness, making it ideal for applications such as lens-making for telescopes and other optical systems. \citet{frumkin2021fluidic, elgarisi2021fabrication} pioneered the fabrication of lenses by immersing a rigid frame in a liquid bath, so that the polymer injected into the frame settles into a rich family of geometries. \citet{luria2023fluidic, luria2024shaping} carried this into microgravity by creating liquid lenses and mirrors in parabolic flights for space telescopes whose size was unconstrained by launch-fairing payload volume. \citet{elgarisi2025fabrication} used the method for wearable optics, fabricating industry-standard prescription eyeglasses in minutes without any machining or polishing. Our work uses the foundational principles of fluidic shaping fabrication, and adds a computational layer to it enabling the creation of arbitrary 3D structures.  

\subsection{Paneling and Segmentation of Freeform Surfaces}
\label{subsec:rw_seg}

Approximating a target shape with a restricted vocabulary of manufacturable
primitives trades exact fidelity for a substantial reduction in fabrication
cost and complexity. In this work, we use piecewise CMC patches as that vocabulary. The closest precedent is
primitive-fitting segmentation. Variational Shape
Approximation~\cite{cohensteiner2004variational} alternates between partitioning
faces and refitting planar proxies under an $L^2$ or normal-based $L^{2,1}$
metric (cf.\ our normal-deviation criterion, Section~\ref{subsec:error}). Later
work added spheres, cylinders and blends~\cite{wu2005structure} and general
quadrics~\cite{yan2006quadric}. Bottom-up methods instead contract dual-graph
edges in order of increasing merge cost~\cite{garland2001hierarchical}, scored
by primitive fits in \citet{attene2006hierarchical}. Our graph contraction based segmentation 
in Section \ref{sec:segmentation} also follows this scheme (see \citet{shamir2008survey} for a survey). 
Although planes, spheres and cylinders are CMC, these methods fit each region a free proxy with
its own parameters, whereas our patches share one $H^*$ and are determined by
their seams, which are physical boundary conditions. Each candidate merge
therefore requires a boundary-value solve (Section~\ref{subsec:cmc_evo}) rather
than a closed-form fit.

Developable segmentation targets fabrication from bent sheet material.
D-Charts~\cite{julius2005d} grows segments from seed triangles until a
developability criterion fails; the same structure as ours with a different
test. \citet{wang2008computing} segment by flattening distortion, and
\citet{zhao2022developability} build on discrete
developability~\cite{stein2018developability}. Other work reduces cost through
reuse: panels sharing molds~\cite{eigensatz2010paneling}, tiles drawn from few
distinct shapes~\cite{fu2010kset, singh2010triangle}, Weingarten surfaces whose
curvature relation induces approximate congruences~\cite{pellis2021computational},
and cladding along geodesic patterns from straight
stock~\cite{pottmann2010geodesic}. \citet{hayashi2023mean}
generate piecewise CMC surfaces by mean curvature flow, but solve the forward
problem for a given patch layout with free internal boundaries. We solve the
inverse, recovering the layout from a target shape with seams fixed by a rigid
frame.

\section{Background on CMC Surfaces and Interfacial Tension Physics}
\label{sec:simulation}

In this section, we characterize CMC surfaces from a geometric and physical perspective. 
We first pose the problem of finding CMC surfaces variationally, as the search for a critical point of an area--volume energy (Section \ref{subsec:cmc_energy}). 
We then show that the equilibrium condition this energy produces is exactly the static force balance of a thin interface under surface 
tension and a pressure jump (Section \ref{subsec:phys_conn}), thus demonstrating equivalence between the geometric and physical formulation.

\subsection{Geometric Modeling of CMC Surfaces}
\label{subsec:cmc_energy}

Let $\bm{X} \in \mathbb{R}^{3n}$ denote the vertex set of a triangle mesh with a fixed boundary, and let $A(\bm{X})$ and $V(\bm{X})$ be its discrete surface area and signed enclosed volume. It is well known \cite{brakke1992surface, pan2012robust, polthier2002discrete} that discrete CMC surfaces arise as the critical points of area under a volume constraint, i.e.\ of the functional
\begin{equation}
    E(\bm{X}) = A(\bm{X}) + \lambda V(\bm{X}),
    \label{eq:energy_lambda}
\end{equation}
where the multiplier $\lambda$ plays the role of a prescribed pressure. Setting $\lambda = -2H^*$ in Equation~\eqref{eq:energy_lambda} gives the energy whose critical points are the CMC surfaces of prescribed target curvature $H^*$,
\begin{equation}
    E(\bm{X}) = A(\bm{X}) - 2H^* V(\bm{X}).
    \label{eq:energy_H}
\end{equation}
We show a simple proof of this in Appendix~\ref{app:crit_cmc}.

We emphasize that we seek a \emph{critical point} of $E$, and not just a minimizer. Consider the major and the minor cap of a sphere. Both share the same boundary circle and both have the same constant mean curvature, but only the minor cap is a local minimizer of $E$; the major cap is a saddle point. Restricting the solve to only minimizers would therefore discard such equally valid solutions. A simple one-dimensional example of this energy for the sphere caps is shown in Fig. \ref{fig:cap_graph}.

\begin{figure}[!t]
  \centering
  \includegraphics[width=0.9\columnwidth]{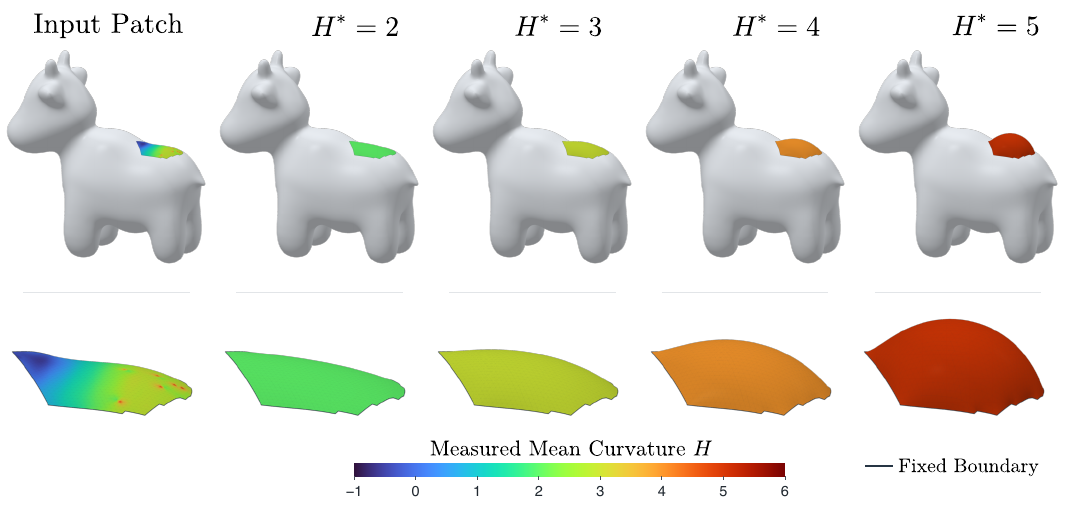}
  \caption{Evolution of a simple patch sampled on Spot after running the CMC solver of Section \ref{subsec:cmc_evo}. We sweep a small family of CMC surfaces of increasing target curvature $H^*$. }
  \label{fig:evo_spot}
\end{figure}

\subsection{Physical Connection: Surface Tension, Pressure and CMC Equilibrium}
\label{subsec:phys_conn}

The critical points of the energy of Equation~\eqref{eq:energy_H} are exactly the static equilibria of a thin interface driven by surface tension and a pressure jump --- the governing physics of the fluidic shaping process (Fig. \ref{fig:pressure_del}). Surface tension acts as an area-minimizing force, pulling the interface inward along $-2\alpha H \bm{n}$, where $\alpha$ is the interfacial tension coefficient. Left unopposed, it would collapse the surface entirely, but the incompressible polymer fluid resists, pushing back with a force proportional to the pressure excess $\Delta p$ across the interface (see Fig. \ref{fig:pressure_del}). We can obtain a static equilibrium between the two by setting the acceleration term in the dynamics model proposed by \citet{ishida2017hgf} to 0, 
\begin{equation}
    \alpha\,\frac{\partial A(x)}{\partial x} = \Delta p \, \bm{n}(x),
    \label{eq:equilibrium_area_grad}
\end{equation}
The stationarity condition $\nabla E(\bm{X}) = 0$ of Section~\ref{subsec:cmc_energy} reduces to exactly this force balance, with the prescribed curvature $H^*$ and the pressure jump related through the Young--Laplace law,
\begin{equation}
  \Delta p = 2\alpha H^*.
  \label{eq:yl}
\end{equation}
The purely geometric formulation of Section~\ref{subsec:cmc_energy} and the static equilibrium of a surface-tension driven interface are therefore equivalent. We defer the derivation to Appendix~\ref{app:phys_conn}.

\begin{figure}[!t]
  \centering
  \includegraphics[width=\columnwidth]{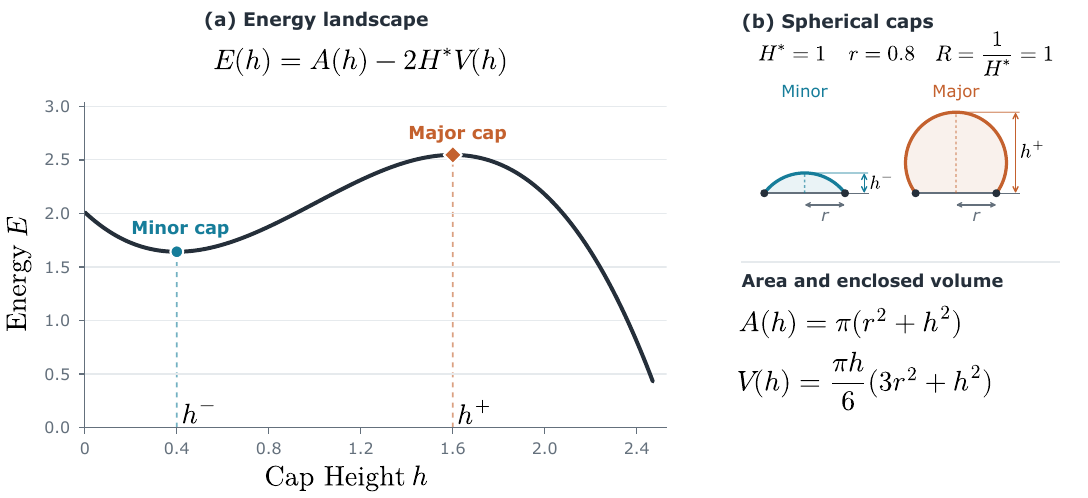}
  \caption{Energy landscape of spherical caps parameterized by their height from the base. The base is the boundary ring which is fixed and has radius $r=0.8$ in this example. In a one-dimensional setting, the minor cap is a local minimizer of $E$ and the major cap is a local maximum. In higher dimensions, the major cap becomes a saddle point of $E$. }
  \label{fig:cap_graph}
\end{figure}

\begin{figure}[!t]
  \centering
  \includegraphics[width=\columnwidth]{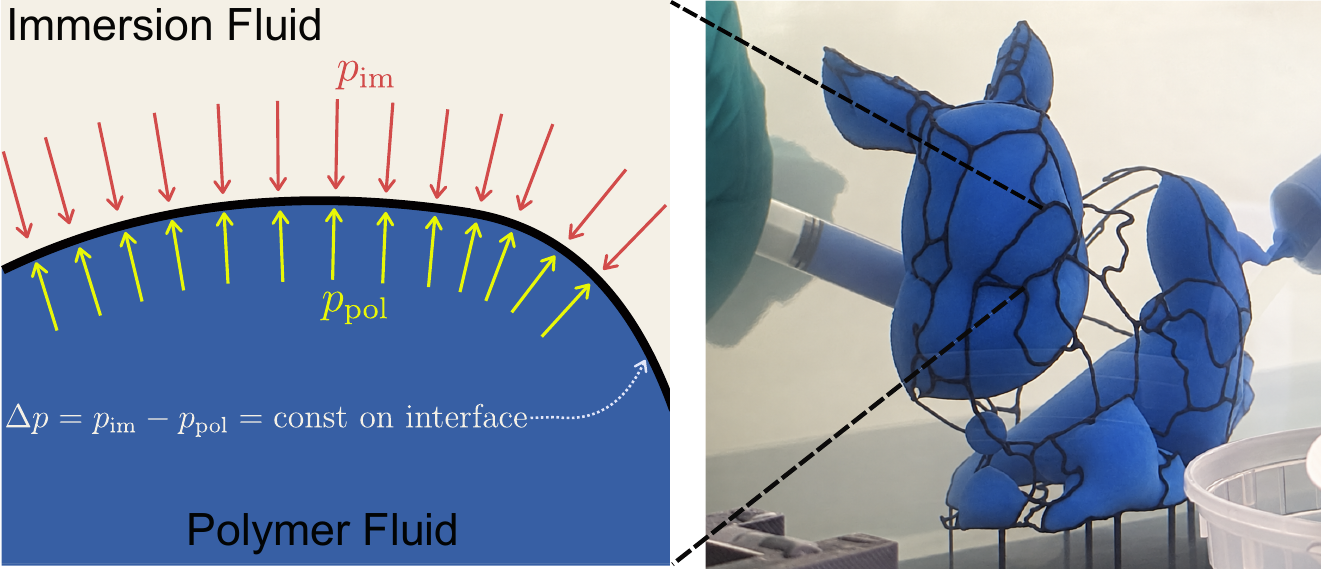}
  \caption{Cross sectional view of the fluidic shaping mechanism. Difference in pressure at the interface of the two fluids $\Delta p$ helps shape the interface. Under neutral buoyancy, $\Delta p$ is constant over the interface and each segment becomes a CMC surface. }
  \label{fig:pressure_del}
\end{figure}

When several polymer segments form within the wireframe but remain in communication through a common fluid body, they share one interior pressure and one ambient bath pressure, and hence a single pressure jump $\Delta p$. 
This follows from the neutral buoyancy of fluidic shaping: because the polymer and the immersion fluid have equal densities, hydrostatic pressure on both sides of the interface is equal, and the pressure jump is therefore determined only due to the mean curvature of the interface.
By Equation~\eqref{eq:yl}, every segment is therefore a CMC patch and all of them share the same target mean curvature $H^*$ at equilibrium. 
This common value of $H^*$ is the target curvature we prescribe in Equation~\eqref{eq:energy_H}.

\section{Mesh Segmentation}
\label{sec:segmentation}

In this section, we describe in detail how a CMC evolution procedure can be leveraged to compute a mesh segmentation. 
The seams of this segmentation are used to extract a thin frame that is 3D printed, which helps shape the polymer fluid into the target geometry. In
Section \ref{subsec:rep_mc}, we show how a suitable target mean curvature $H^*$ can be chosen for the input geometry. In Section \ref{subsec:cmc_evo},
we describe the process to compute a CMC surface with MC $H^*$ for a given patch on the target geometry. We leverage these capabilities 
in Section \ref{subsec:merging} to compute the P-CMC mesh decomposition in a ground-up, geometrically-driven merging approach. Finally, in Section \ref{subsec:merge_cons}, we detail some constraints 
the segmentation must satisfy in order to produce a high quality P-CMC decomposition. 

\subsection{Representative Mean Curvature}
\label{subsec:rep_mc}

Before we begin describing the segmentation routine, we need to pick a good representative mean curvature $H^*$ for the segmentation. As discussed earlier, all segments share the same mean curvature at equilibrium (Section~\ref{subsec:phys_conn}).

Intuitively, it makes sense to pick the mean curvature that is most dominant on the mesh. This helps ensure that the resulting segments cover larger areas of the mesh, thereby reducing the amount of supporting frame that must be 3D printed and maximizing the portion of the surface that remains smooth.

To find this value, we compute the mean curvature at every vertex and discard the 2\% of vertices with the lowest and highest values to remove outliers. We then build a smooth, area-weighted distribution of the remaining values using Gaussian Kernel Density Estimation (KDE)~\cite{rosenblatt1956remarks,parzen1962estimation}. We place a Gaussian centered at each vertex's mean curvature, weight it by that vertex's share of the total surface area, and add them together. The result is a Probability Density Function (PDF), and its height at any value $H$ reflects how much of the surface has mean curvature close to $H$.
All Gaussians share the same width (standard deviation) $h$, which we set using Silverman's rule of thumb~\cite{silverman1986density}: $h = 1.06\,\hat{\sigma}\,n^{-1/5}$, where $\hat{\sigma}$ is the area-weighted standard deviation of the vertex mean curvatures and $n$ is the number of vertices. 

Once the PDF is obtained, we evaluate the density at 1000 evenly spaced values spanning the range of mean curvatures, and pick the value with the highest density as $H^*$.

\begin{figure}[!t]
  \centering
  \includegraphics[width=\columnwidth]{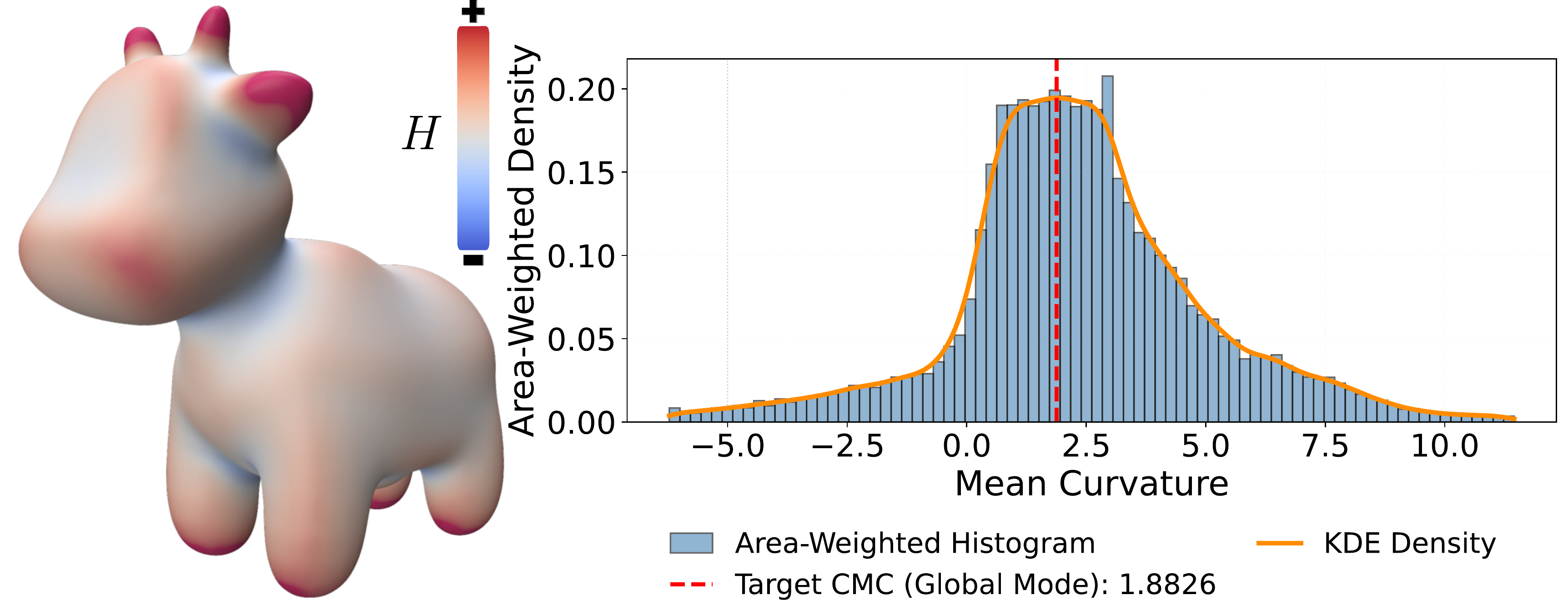}
  \caption{KDE fitting for MC estimation for Spot. Left: Original mean curvature plot. Right: KDE fitting used to compute peak MC value. This peak value is used as the representative mean curvature $H^*$ for the segmentation. The histogram is only for visualization purposes, and is not used in the computation.}
  \label{fig:estimate}
\end{figure}

\subsection{Evolution into CMC Surfaces}

\label{subsec:cmc_evo}

Given an input surface with a fixed boundary, we describe how it is evolved into a CMC surface, i.e.\ how we find a nearby critical point of $E$. We saw how critical points of $E$ are precisely the CMC surfaces of target curvature $H^*$. The simplest approach to finding a critical point is to descend on the energy, but this is only guaranteed to converge to a local minimizer. Certain saddle points, which are equally valid solutions (Fig. \ref{fig:cap_graph}), are therefore lost.

We root-find instead, in two phases. A derivative-free outer search first brings the surface to the prescribed curvature within a loose tolerance. Newton's method on the curvature residual then takes over and converges quadratically, indifferent to whether the critical point is a minimum or a saddle. The goal of the first phase is to place the surface close enough for the second to converge reliably.

\paragraph{Reaching Newton's basin.}
Following \citet{pan2012robust}, we do not prescribe the curvature directly. Their energy
\begin{equation}
    F(\bm{X}; t) = F_{\mathrm{area}}(\bm{X}) + t \, F_{\mathrm{vol}}(\bm{X})
    \label{eq:pan_energy}
\end{equation}
weights the volume term by a scalar $t$ that acts as a force rather than a curvature. The central contribution of their work is that minimizing $F$ over the vertex positions $\bm{X}$ is stable across a far wider range of surfaces than prescribing $H^*$ outright. The price we pay is that curvature becomes an output: the minimizer is a CMC surface, but its mean curvature $H(t)$ is now a function of $t$. It is also a physically \emph{stable} one --- we show in Appendix~\ref{app:stability} that a local minimizer of $F$ admits no volume-preserving deformation that decreases area.

Finding the CMC surface is therefore a scalar root-finding problem on $t$. We define the residual as:
\begin{equation}
    g(t) = H(t) - H^*
    \label{eq:residual_t}
\end{equation}
and find $t^\ast$ with $g(t^\ast) \approx 0$. Every evaluation of $g$ runs a full inner minimization of Equation~\eqref{eq:pan_energy} and extracts the area-weighted mean of its interior vertex mean curvatures from the resulting surface. We want the outer search to run as few of these evaluations as possible. Newton's method is unavailable here, since $dH/dt$ would require differentiating through an $\arg\min$ and a finite-difference estimate would cost a full extra minimization per step. We use a derivative-free method instead, built only from the $(t, g)$ pairs we already compute.

\begin{figure}[!t]
  \centering
  \includegraphics[width=0.85\columnwidth]{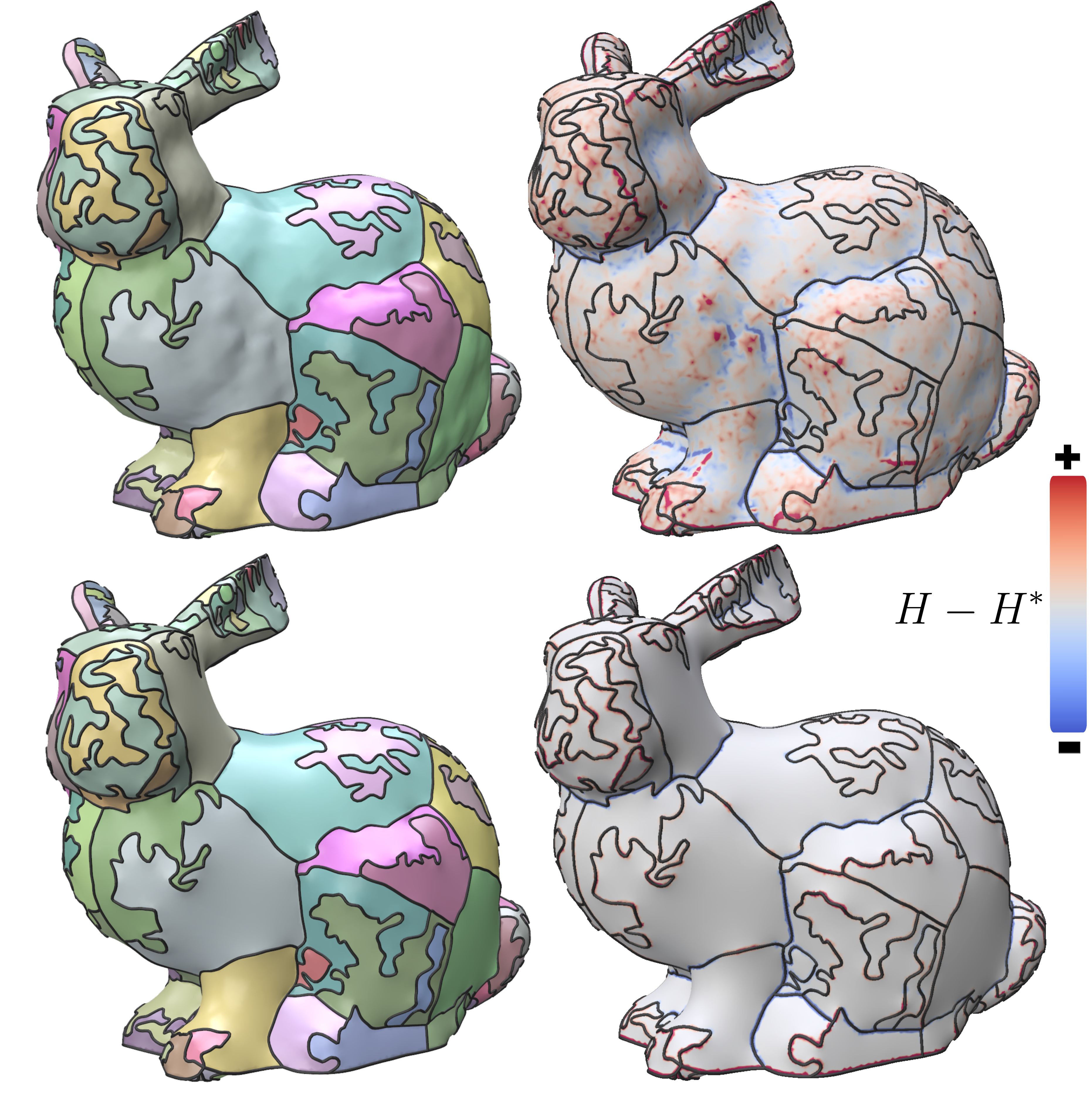}
  \caption{Evolution of the bunny mesh (top row) into a piecewise CMC surface (bottom row) after running CMC evolution of Section \ref{subsec:cmc_evo}. Target MC, $H^*=0.8$. Seams are shown in black and are fixed during the evolution process. We describe the full segmentation procedure in Section \ref{sec:segmentation}.}
  \label{fig:evo_cmc_full_seg}
\end{figure}

The search runs in one of two regimes, selected by whether the evaluations so far bracket the root. When two of them straddle zero, we take an Illinois step \cite{dowell1971modified} giving us superlinear convergence. The bracket is preserved at every step, so the root can never be lost. When no bracket has been found yet, we fall back to a secant step using the two most recent evaluations, clamped by a trust radius so that a badly extrapolated chord cannot leave the branch we are solving on. Both regimes cost one $g$ evaluation per iteration. Algorithm~\ref{alg:two_path} in Appendix~\ref{app:root_search} gives the full search.

\paragraph{Newton refinement.}
The outer search gives us a surface whose mean curvature is already close to $H^*$. Newton's method enables us to fully converge, working directly with the curvature residual. The correction we solve for is a normal displacement of the surface: a scalar field $\varphi$ that vanishes on the fixed boundary and moves the vertices by
\begin{equation}
    \bm{X} \leftarrow \bm{X} + \varphi \, \bm{n}.
    \label{eq:normal_offset}
\end{equation}
Sliding a vertex tangentially does not change the surface, and so does not change its curvature. Such directions lie in the kernel of the Jacobian $\mathcal{J}$ of Equation~\eqref{eq:jacobi_op}, and would make the linearized system singular. Restricting to normal displacements removes that null space and reduces the unknowns from $3n$ to one per interior vertex.
Newton's method computes a $\varphi$ that drives the pointwise residual
\begin{equation}
    r = \bm{H} - H^*
    \label{eq:newton_residual}
\end{equation}
to zero at every interior vertex, to first order in $\varphi$. Because $H^*$ is a constant, linearizing $r$ means linearizing the mean curvature, whose first variation under a normal displacement is the classical Jacobi operator \cite{polthier2002discrete, brubaker2018continuation},
\begin{equation}
    \delta H[\varphi] = -\tfrac{1}{2}\left(\Delta \varphi + |A|^2 \varphi\right) = -\tfrac{1}{2}\mathcal{J}\varphi,
    \qquad
    \mathcal{J} = \Delta + |A|^2,
    \label{eq:jacobi_op}
\end{equation}
where $\Delta$ is the Laplace--Beltrami operator and $|A|^2 = 4H^2 - 2K$ is the squared norm of the second fundamental form. Setting the linearized residual to zero, $r + \delta H[\varphi] = 0$, gives the Newton step
\begin{equation}
    \mathcal{J}\varphi = 2r \ \text{ on } \Sigma,
    \qquad
    \varphi = 0 \ \text{ on } \partial\Sigma.
    \label{eq:newton_continuous}
\end{equation}

For discretization, we use the cotan Laplacian matrix $\bm{L}$ and the diagonal mass matrix $\bm{M}$, so that $\Delta \approx -\bm{M}^{-1}\bm{L}$ and $\mathcal{J} \approx -\bm{M}^{-1}\bm{L} + \mathrm{diag}(|A|^2)$. Multiplying through by $\bm{M}$ symmetrizes the system. Writing
\begin{equation}
    \bm{S} = \bm{L} - \mathrm{diag}\!\left(\bm{M} \odot |A|^2\right),
    \label{eq:jacobi_matrix}
\end{equation}
with $\odot$ the elementwise product, the Newton step becomes the sparse symmetric system
\begin{equation}
    \bm{S}_{II}\, \varphi_I = -2\left(\bm{M} \odot r\right)_I,
    \qquad
    \varphi_B = 0,
    \label{eq:newton_system}
\end{equation}
where $I$ and $B$ index the interior and boundary vertices. Vertex curvatures come from $H_i = \tfrac{1}{2}\langle (\bm{M}^{-1}\bm{L}\bm{X})_i, \bm{n}_i \rangle$, and $K_i$ from the angle defect.

\begin{figure}[!t]
  \centering
  \includegraphics[width=\columnwidth]{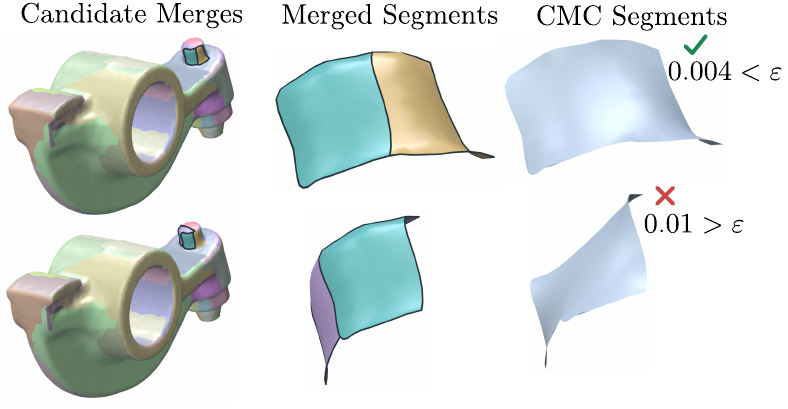}
  \caption{Example of a good and bad merge. Top - Merged segments and their CMC counterpart stay within the error threshold, indicating a successful merge. Bottom - A failed merge operation where the resulting CMC surface deviates significantly from the target surface. }
  \label{fig:good_bad_merge}
\end{figure}

$\bm{S}$ is an approximation to the true Jacobian of the residual with respect to all $3n$ coordinates, since it accounts only for normal motion and holds the operators fixed while differentiating. Each Newton iteration refreshes the operators, since the geometry of the surface changes with each iteration. We assemble $\bm{L}$, $\bm{M}$, $\bm{n}$, $H$, $K$ and $|A|^2$ at the current positions, form the residual, solve Equation~\eqref{eq:newton_system} for $\varphi$, and take the largest damping factor $s \in \{1, \tfrac{1}{2}, \tfrac{1}{4}, \dots\}$ for which $\bm{X} + s\varphi\bm{n}$ reduces $\|r\|$. The line search is on the residual and never on $E$, since backtracking on the energy would reintroduce the descent bias we are trying to avoid. A handful of steps suffices, with the residual falling quadratically since we are already close to the solution.

\paragraph{Convergence.}
We terminate once $\|r\|_\infty$ on the interior vertices falls below a tolerance, typically set to $10^{-6}$. If the residual fluctuates or fails to converge smoothly, 
we take this as an empirical indication that no CMC surface exists for the specified target mean curvature and the given boundary. 

Fig. \ref{fig:evo_spot} shows simple examples of the evolution process and Fig. \ref{fig:simple_seg} shows segmentation results on simple geometries produced using this CMC evolution procedure. The results of this procedure on the fully segmented Stanford Bunny can be seen in Fig. \ref{fig:evo_cmc_full_seg}.

\begin{figure}[!t]
  \centering
  \includegraphics[width=\columnwidth]{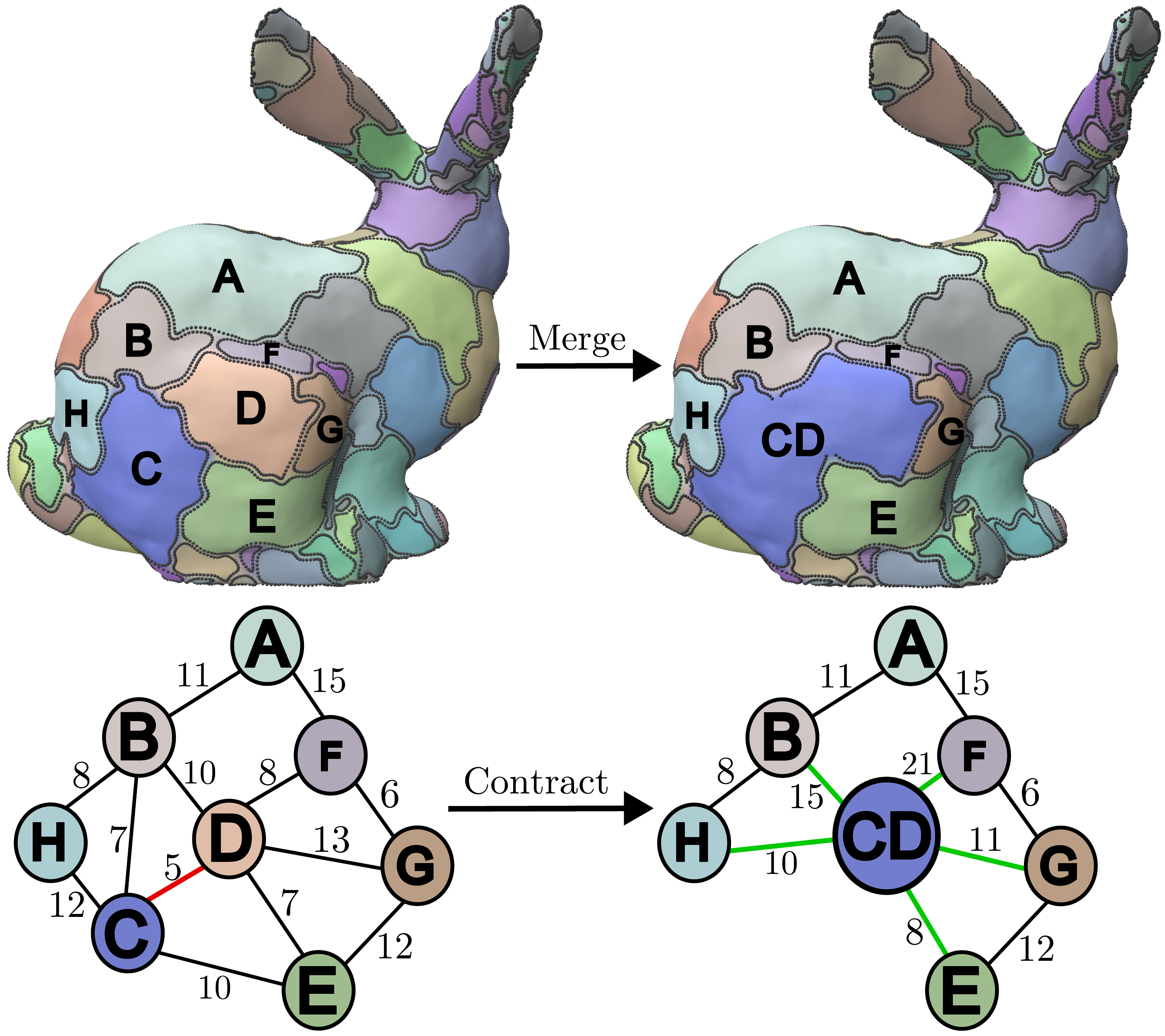}
  \caption{Graph contraction for merging. The example shows how the graph encodes the segmentation. Top: Segments \textbf{C} and \textbf{D} are merged together, since edge \textbf{CD} in red has the lowest cost. Bottom: Nodes \textbf{C} and \textbf{D} are contracted in the graph to reflect the merge above. The costs of the new edges formed also need to be updated, shown in green. }
  \label{fig:contract_seg}
\end{figure}

\subsection{Segmentation - Merging through Graph Contractions}
\label{subsec:merging}

\looseness=1
Now that we have a way to compute a target $H^*$, and a CMC evolution procedure in place to tell us how much a potential segment on the surface deviates from its CMC counterpart, we can proceed with the description of the merging process for segmentation.
We approach the segmentation problem through a graph contraction lens, where segments grow from a single element (mesh face).

\looseness=1
Initially, every element is its own segment. We encode the segmentation as a graph whose nodes are elements and whose edges join adjacent elements, denoting potential expansion candidates for a segment. 
See Fig. \ref{fig:contract_seg} for an example. Each edge stores the cost of merging the two segments it connects, computed as the error between the evolved/fitted CMC segment and the original 
geometry. Similar to QEM \cite{garland1997surface}, we maintain a priority queue of these costs and contract the cheapest pair every iteration. We compute this cost of merging two segments
as follows:

\subsubsection{Surface Approximation Error}
\label{subsec:error}

We score the cost of merging two segments by the root mean square error (RMSE) between the merged segment and the original target surface. Let $\bm{x}_1, \dots, \bm{x}_m$ be the vertices of the combined segment on the target mesh and $\tilde{\bm{x}}_1, \dots, \tilde{\bm{x}}_m$ their positions after the CMC evolution. The evolution usually leaves the mesh topology unchanged, so we measure the error between corresponding points,
\begin{equation}
    \mathrm{RMSE} = \sqrt{\frac{1}{m}\sum_{i=1}^{m} \bigl\| \tilde{\bm{x}}_i - \bm{x}_i \bigr\|^2 }.
    \label{eq:rmse}
\end{equation}
If the topology does change, for instance through remeshing, we instead take $\tilde{\bm{x}}_i$ to be the closest point to $\bm{x}_i$ on the evolved patch.

\begin{figure}[!t]
  \centering
  \includegraphics[width=\columnwidth]{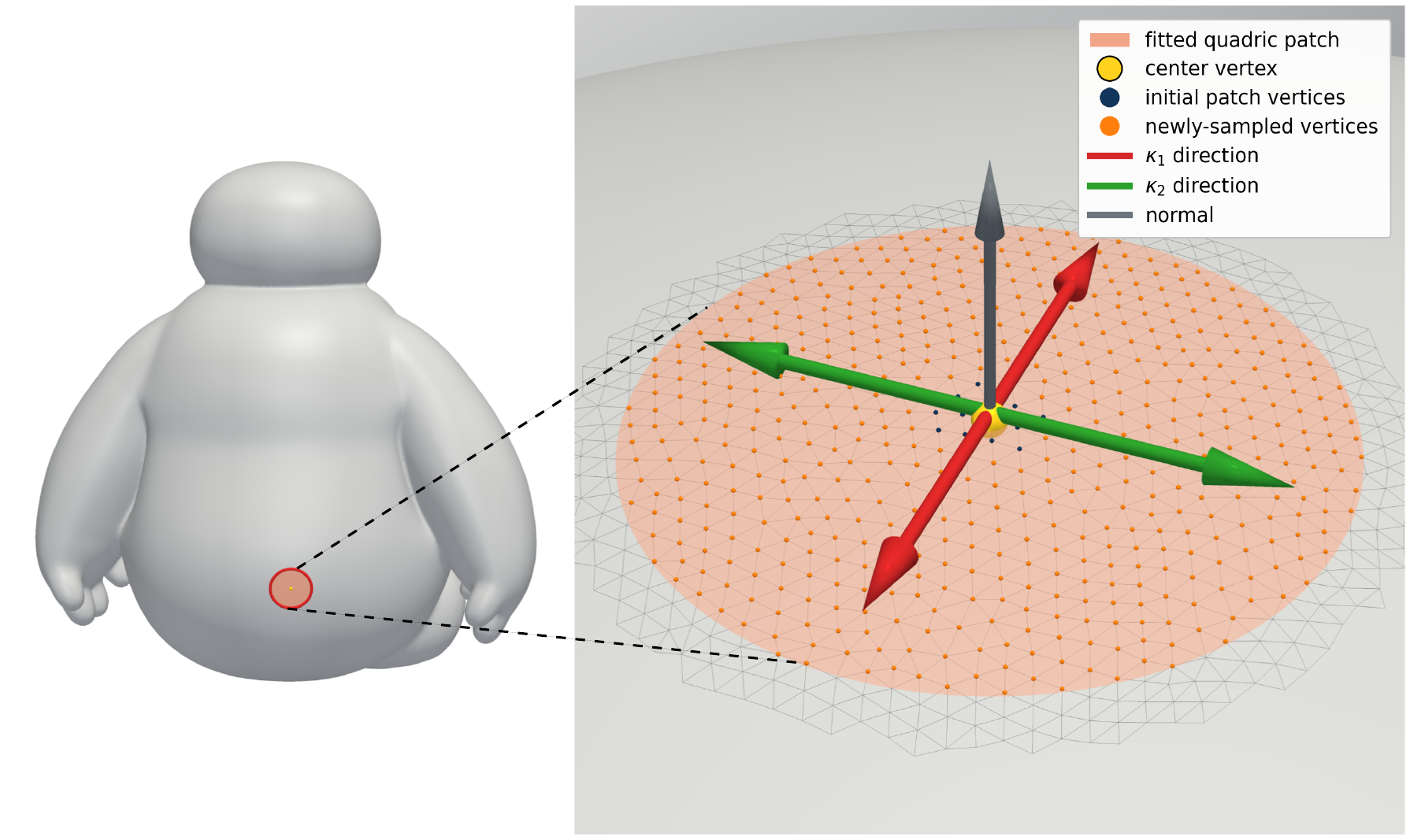}
  \caption{Quadric Fitting on Baymax. The target MC $H^*=0.06$ for this example with $\kappa_1 \approx \kappa_2$ at the center vertex (shown in yellow). The mean curvature at the center is $H_C =\frac{\kappa_1+\kappa_2}{2} = H^* = 0.06$. }
  \label{fig:fitting}
\end{figure}

Each contraction forms a merged node containing multiple elements. We keep contracting until the error between a segment's CMC approximation and its original geometry exceeds a user defined threshold (Fig. \ref{fig:good_bad_merge}), typically $\varepsilon$\% of the diagonal of the mesh bounding box (set to $1\%$ for most of our experiments). 

\subsubsection{Handling Small Segments with MC Constrained Quadrics}
\label{subsec:quadrics}

\begin{wrapfigure}{r}[20pt]{0.40\columnwidth}
  \hspace*{-12pt}\includegraphics[width=0.40\columnwidth]{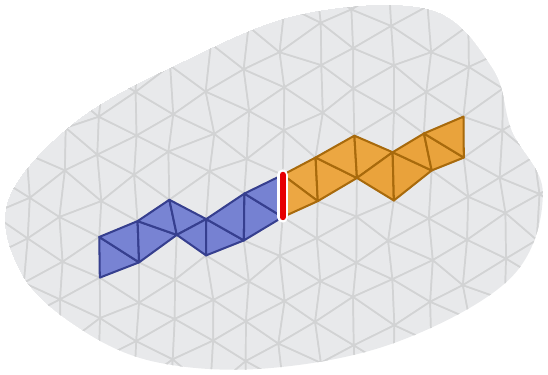}
\end{wrapfigure}
The greedy nature of the algorithm described above can exploit the cost function in an undesirable fashion. Consider two segments whose union would be composed of a strip of triangles (inset).
Every vertex in this strip sits on the boundary, and it contains no interior vertices. The evolution of Section \ref{subsec:cmc_evo} pins boundary vertices, so there is nothing left to solve for and the patch is unchanged. 
The cost of contracting these segments is therefore 0. Since we always contract the cheapest edge, these segments merge first, and we get long thin strips that continue to grow without bounds. 
This is where the quadric fitting approach we describe next helps. It doesn't reward thin meandering strips and promotes the creation of relatively rounder segments by sampling new elements 
uniformly during the initial phase of the segmentation. Once the segments have grown sufficiently and have enough resolution, we switch to the CMC evolution procedure described in Section \ref{subsec:cmc_evo}.

We characterize a small segment as one having fewer than 50 elements. Until we have enough elements for a robust CMC evolution, we sample points uniformly around the segment within a ball of a small radius proportional to the average mesh edge length, similar to \cite{panozzo2010efficient}. 
The quadric patch is rooted at the centroid of the vertices in the segment. 
The centroid is constrained to have mean curvature equal to the prescribed MC $H^*$ at that point. Fig. \ref{fig:fitting} illustrates this process with an example. 
This helps serve as a useful proxy for a CMC patch for the initial part of the segmentation.
As the segments grow larger, the quadric approximation becomes less accurate, so we switch to the evolution of Section \ref{subsec:cmc_evo}, which by then has enough 
resolution to work with.

\begin{figure}[!t]
  \centering
  \includegraphics[width=\columnwidth]{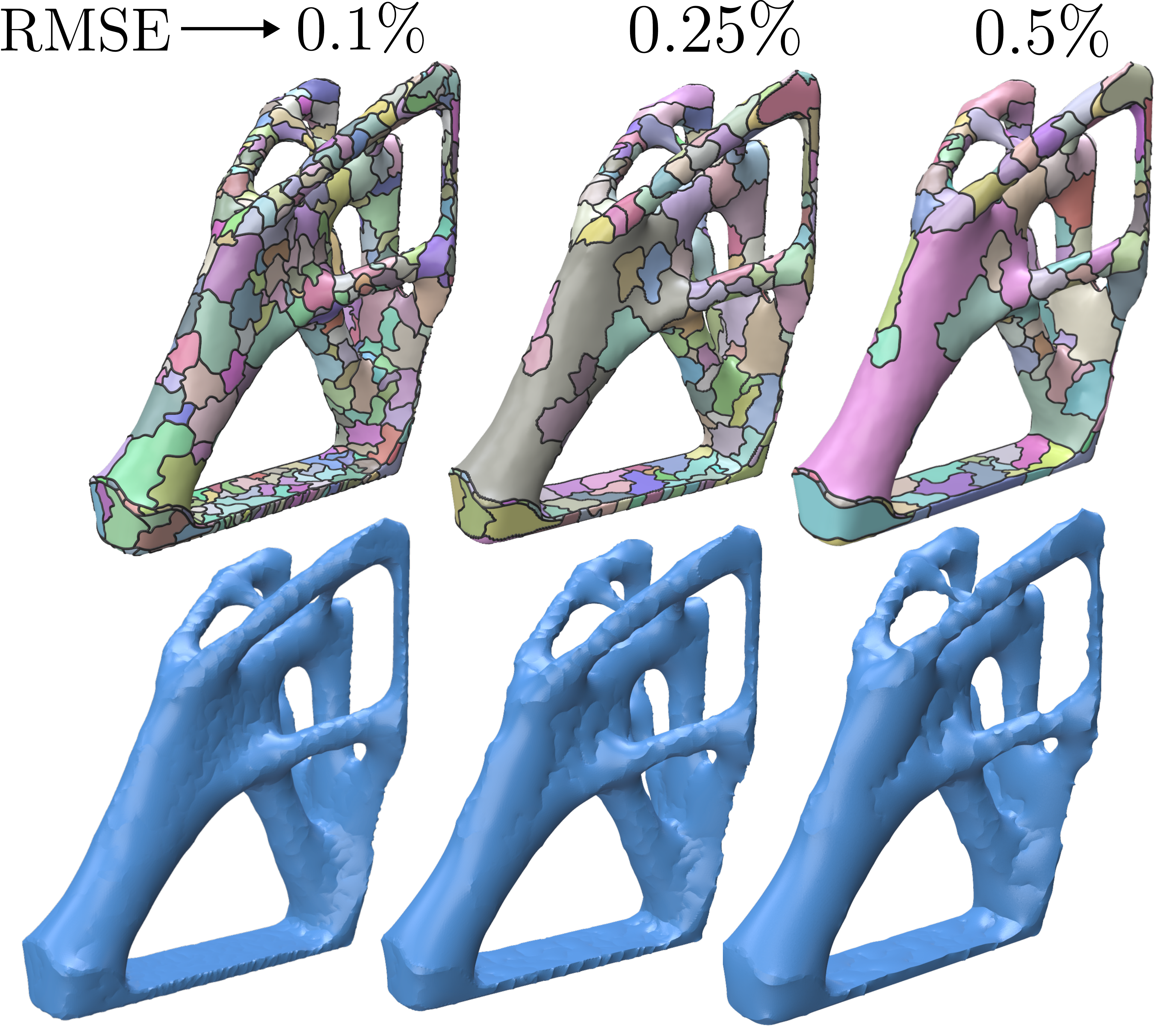}
  \caption{Impact of tuning plain RMSE for error threshold on shape approximation. Left to Right: 882, 381 and 231 segments. Target MC for all is $H^*=0.02$. Top Row: Segmentation results for various RMSE thresholds. Bottom Row: Evolved piecewise CMC structures. }
  \label{fig:rmse_comp}
\end{figure}

\paragraph{Constrained Least Squares Fitting for Quadrics}
Given the centroid \(\bm{p}\) of a segment, we build an orthonormal frame \((\bm{t}_1,\bm{t}_2,\bm{n})\) at \(\bm{p}\), where \(\bm{t}_1\) and \(\bm{t}_2\) span the tangent plane and \(\bm{n}\) is computed similar to \cite{panozzo2010efficient} (see blowup of Fig. \ref{fig:fitting}). Expressing nearby points in this frame turns the local surface into a height field over the tangent plane. Small segments have too few vertices of their own to fit against, so we gather samples from a ball of radius twice the average edge length of the mesh.

We model the patch as a quadratic height graph
\begin{equation}
h = w(u,v) = a u^2 + b u v + c v^2 + d u + e v + f,
\label{eq:quadric_model}
\end{equation}
and fit its six coefficients to the samples in a least squares sense, subject to two constraints.

The first constraint makes the centroid a critical point of the graph. Requiring both first derivatives to vanish at the local origin reduces to \(d = 0\) and \(e = 0\) for this model. We call these the apex constraints, even though the critical point can be a saddle.

The second constraint prescribes the curvature. At a critical point of a height graph the mean curvature is \(H = (\frac{\partial^2 w}{\partial u^2} + \frac{\partial^2 w}{\partial v^2})/2\), which for Equation~\eqref{eq:quadric_model} is simply \(a + c\). A target mean curvature \(H^\ast\) is therefore the linear constraint \(a + c = H^\ast\).

Both constraints are linear in the coefficients, so instead of solving a constrained problem we substitute them into Equation~\eqref{eq:quadric_model} directly. Three free coefficients remain, and the fit becomes an ordinary linear least squares solve. Appendix \ref{app:quadric} provides the full derivation. The result is a patch that fits the local geometry and has exactly the prescribed mean curvature at its apex/centroid.

The cost of fitting this quadric is also computed via an error calculation (see Section \ref{subsec:error}) between the fitted patch and the sampled points on the mesh. 
Each sample is paired with the point on the quadric directly above or below it in the local frame, so the RMSE is taken over the height residuals $w(u_j, v_j) - h_j$.

\subsection{Merge Constraints}
\label{subsec:merge_cons}

The wireframes extracted from the seams of the above segmentation process can sometimes be overly complex. 
These are thin intricate structures that are 3D printed and may require substantial support material. We keep the fabrication complexity in check by 
enforcing the following criteria and describe an approach to improve the plain RMSE metric.  

\subsubsection{Segment Manifoldness for Printability}

We prevent the formation of non-manifold segments by setting the cost to create them to $\infty$. This corresponds to setting the edge cost in the graph between nodes whose union would give non-manifold segments to $\infty$.

\begin{figure}[!t]
  \centering
  \includegraphics[width=\columnwidth]{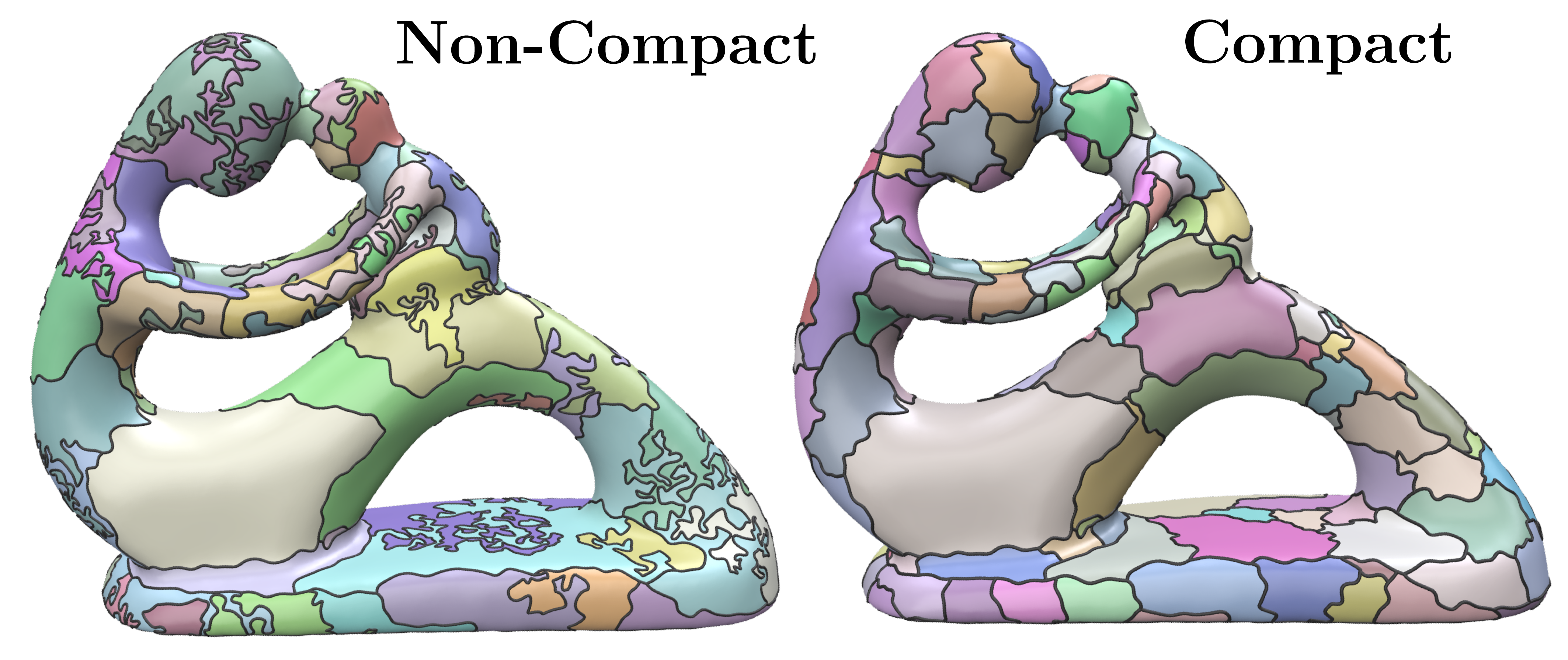}
  \caption{Comparing segment Compactness and non-Compactness. Left: No compactness criteria enforced, segments are allowed to have free-flowing boundaries. Right: Compactness criteria enforced. All segments have a compactness score $\ge$ 0.3. }
  \label{fig:comp_seg}
\end{figure}

\subsubsection{Compactness}
\label{subsubsec:compact}

Enforcing segment compactness helps reduce the area occupied by the extracted frame on the mesh. However, the right balance is important, 
since an overly strong compactness requirement can have the opposite effect. 
We use the inverse of the irregularity measure in \cite{garland2001hierarchical} to quantify segment compactness. 
For planar circular disk shaped segments, this measure approaches 1, indicating high compactness, while for elongated or 
jagged segments, it approaches 0, indicating low compactness. It may exceed 1 in case of non-planar patches, but gives
a practical method for mitigating most complex segments. Figs. \ref{fig:comp_seg} and \ref{fig:bunny_comp} compare the 
compact and non-compact segmentation outcomes. It's important to note that we use this as a threshold to prevent merging
of certain segments, and not as an additional term added to the cost. Our examples typically use a compactness threshold of 0.3. 
In practice, we set the edge weight between two nodes (segments) to be $\infty$ if merging them would create a segment with compactness score less than 0.3. 

\subsubsection{Normal Deviation Threshold}

\begin{figure}[!t]
  \centering
  \includegraphics[width=0.85\columnwidth]{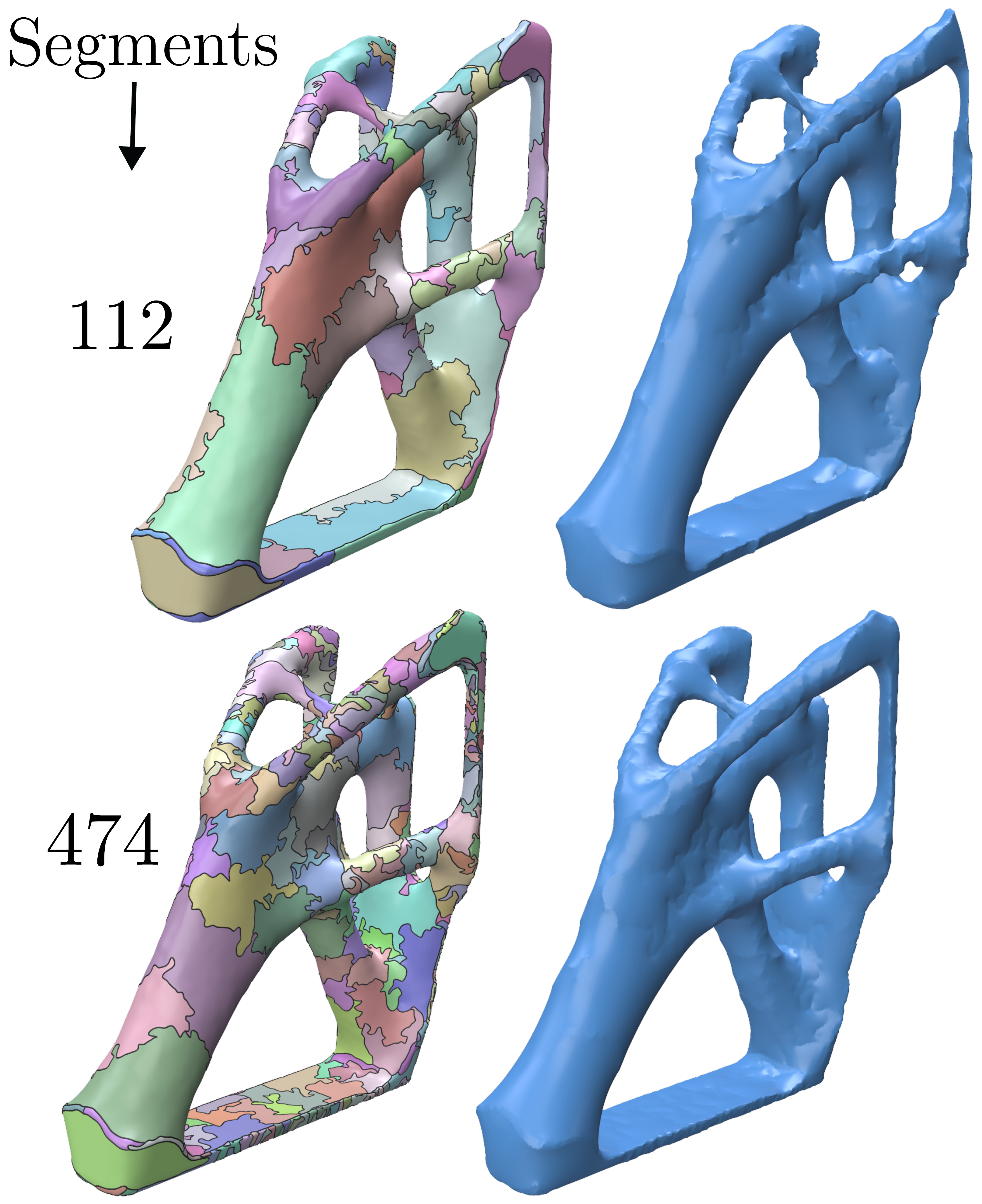}
  \caption{Normal Deviation Threshold helps segmentation adapt to local geometry. Target MC for both examples is $H^*=0.02$. Left Column: Segmentation results. Right Column: Evolved piecewise CMC surfaces. Top Row: Only RMSE threshold set to $1\%$. Bottom Row: RMSE threshold set to $1\%$ + Normal deviation tolerance set to $20^{\circ}$. }
  \label{fig:normal_dev}
\end{figure}

We propose a threshold that helps approximate bulk regions with fewer
segments while capturing sharper regions more faithfully, effectively adapting to
local geometry. The metric takes a normal-deviation threshold angle as input: if any vertex normal on the evolved CMC patch deviates from 
the target by more than this angle, the corresponding merge is disallowed by setting that edge's cost to
$\infty$. Fig. \ref{fig:normal_dev} shows the effect of this threshold on the segmentation.

\section{Implementation Details}

This section describes the lower-level details that make the segmentation algorithm reliable in practice, including the initialization strategy and the constraints we impose on the segmentation. We then show how the segmentation is used to extract the bounding frame, and close with details of the fabrication process.

\begin{figure}[!t]
  \centering
  \includegraphics[width=0.9\columnwidth]{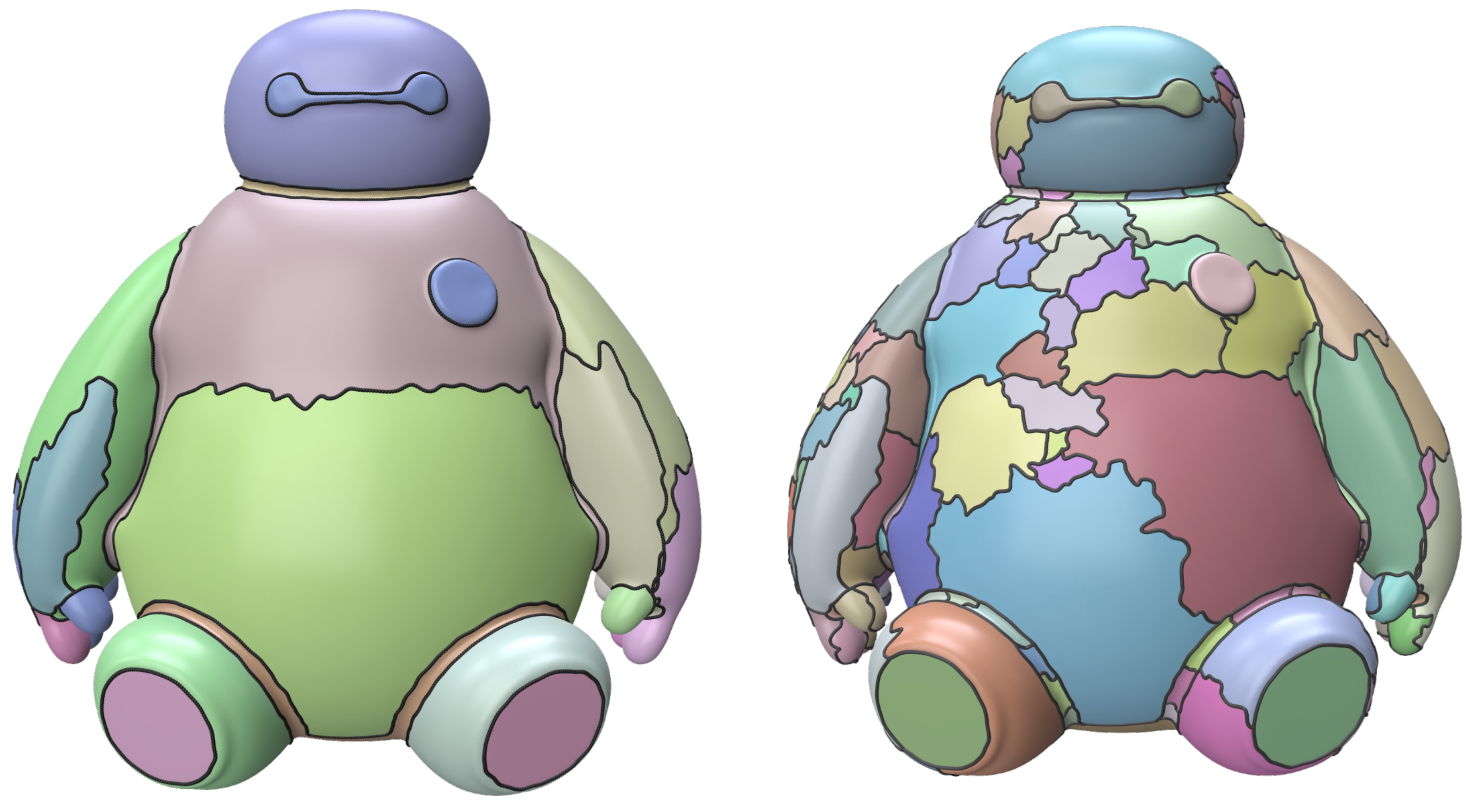}
  \caption{Left: Initial Segmentation. Right: P-CMC segmentation that follows ensures cuts from initial segmentation are preserved. }
  \label{fig:init_seg}
\end{figure}

\subsection{Initial Segmentation for Cut Preservation}

Our segmentation method can preserve cuts from an existing initial segmentation. Such cuts can be vital in regions 
of sharp detail that must not be smoothed over or lost. In Fig. \ref{fig:init_seg}, we initialize with spectral clustering 
and then apply our CMC segmentation. All cuts from the initial segmentation are preserved in the final result. Spectral clustering
was run using an affinity graph penalizing differences between mean curvatures of adjacent vertices \cite{von2007tutorial}. This
promotes cuts in regions of high mean curvature variation.

\subsection{Solid Regions}

Some regions of the structure may need to be 3D printed in full rather than only at the seams between segments, when exact geometric fidelity is required. 
This can be enforced as an additional constraint on the segmentation. One option is to detect such regions 
automatically with a simple criterion such as flatness: any vertex whose mean curvature is very close to zero 
is marked as an isolated vertex in the graph. Isolated vertices cannot be merged with any other vertex during 
the contraction process, so a contiguous patch of them is never absorbed into neighboring segments. Alternatively, 
regions can be specified manually by designating certain segments of an initial segmentation to be fully solidified. 
In both cases, the faces of these regions are added to the strip extracted from the segmentation seams, so they 
are printed together with the frame. See Fig.~\ref{fig:solid} (the back of the bracket) for an example.

\subsection{3D Printing the Frame}

\subsubsection{Seam Smoothing}
We use Taubin Smoothing \cite{taubin1995curve} to smooth wiggly seams. The smoothing helps even out seams with high local curvature fluctuations. 
This helps reduce stress concentrations on the frame and also adds to the aesthetic appeal. 

\subsubsection{Frame Extraction and Solidification}
\begin{figure}[!t]
  \centering
  \includegraphics[width=\columnwidth]{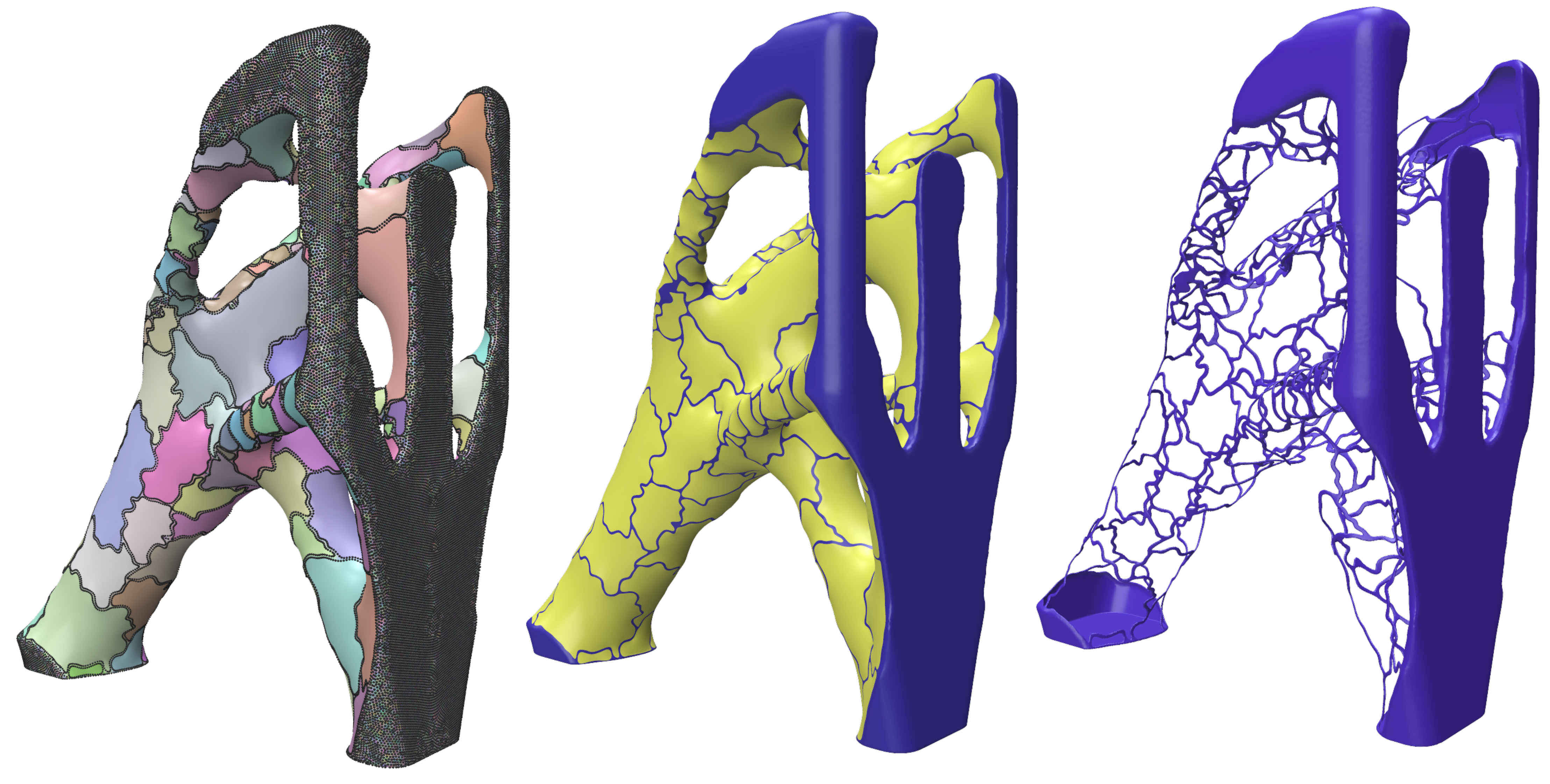}
  \caption{Left to Right: Segmentation, Thin strip extraction and Final 3D printable frame extraction. The back of the bracket is made solid, and will be 3D printed. }
  \label{fig:solid}
\end{figure}

The thin strip extracted from the seams cannot directly be 3D printed because it lacks volume. We create a UDF for the seam network that adds uniform thickness in all directions, and use marching cubes for the final mesh extraction. This guarantees self-intersection-free, 3D printable frames (see Fig. \ref{fig:solid}). 
Because our structures are thin, wireframe-like scaffolds, they require substantial support material to print. We use dissolvable supports so they can be removed easily during post-processing.

\subsection{Fluidic Fabrication}

Once the wireframe scaffold is 3D printed, it serves as a support structure for the fluidic fabrication process. 
We immerse the scaffold into a bath whose density matches that of the liquid polymer. The polymer is then deposited onto 
the scaffold, and interfacial tension between the two fluids causes it to form a smooth, continuous surface that conforms 
to the underlying wireframe. Some liquid polymers self cure, while others can be cured via UV light. 
We refer the reader to \cite{Garrett2026Fluidic} for a more comprehensive description of the fabrication process. 

\begin{figure*}[!t]
\centering
\includegraphics[width=\textwidth]{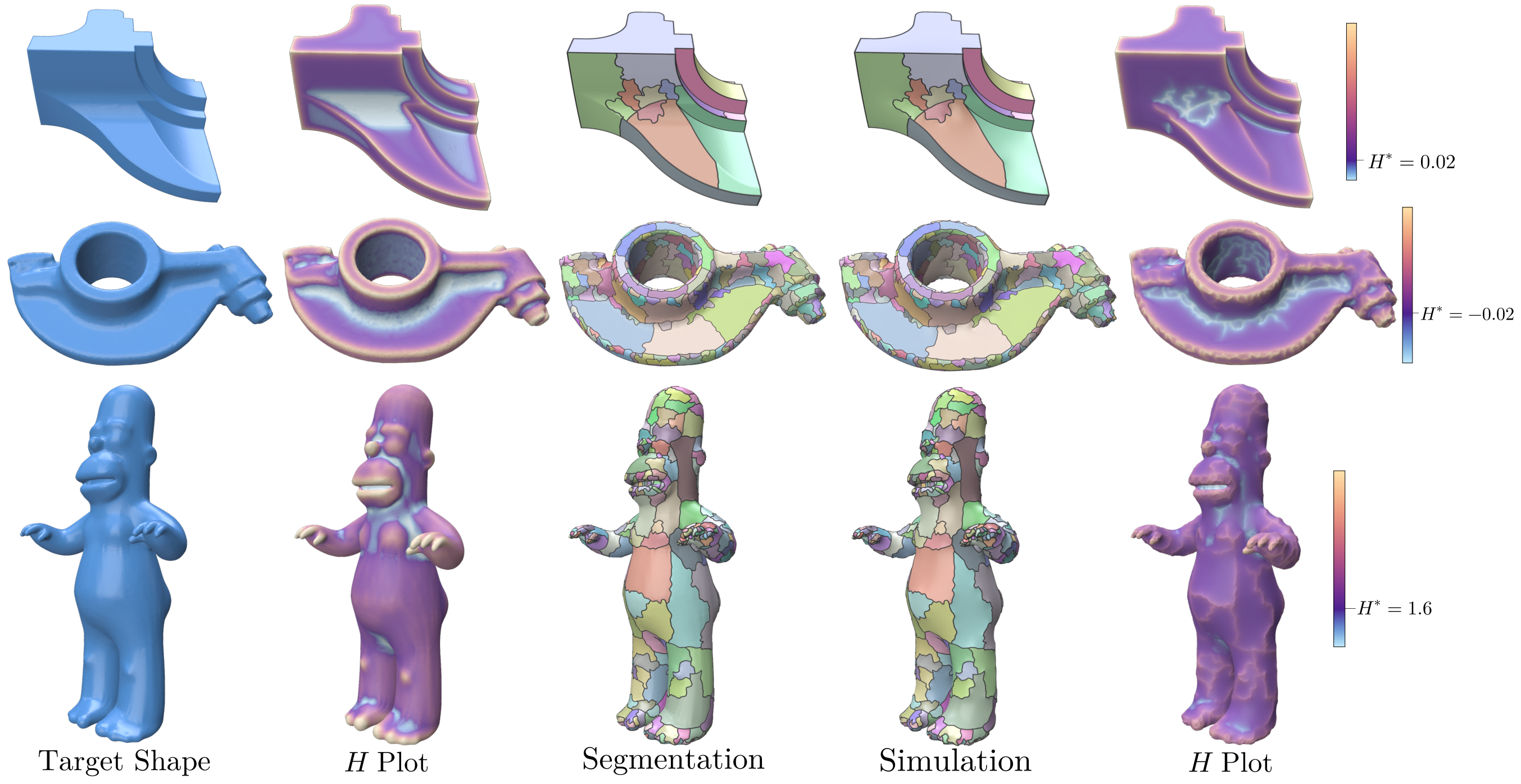}
\caption{Simulated results for various meshes. We use target mean curvatures $H^* = 0.02$, $H^* = -0.02$ and $H^* = 1.68$ for the Fandisk, Rocker and Homer meshes respectively computed from the MC estimator of Section \ref{subsec:rep_mc}.}. 
\label{fig:sim_results}
\end{figure*}

\begin{figure*}[t]
\centering
\includegraphics[width=\textwidth]{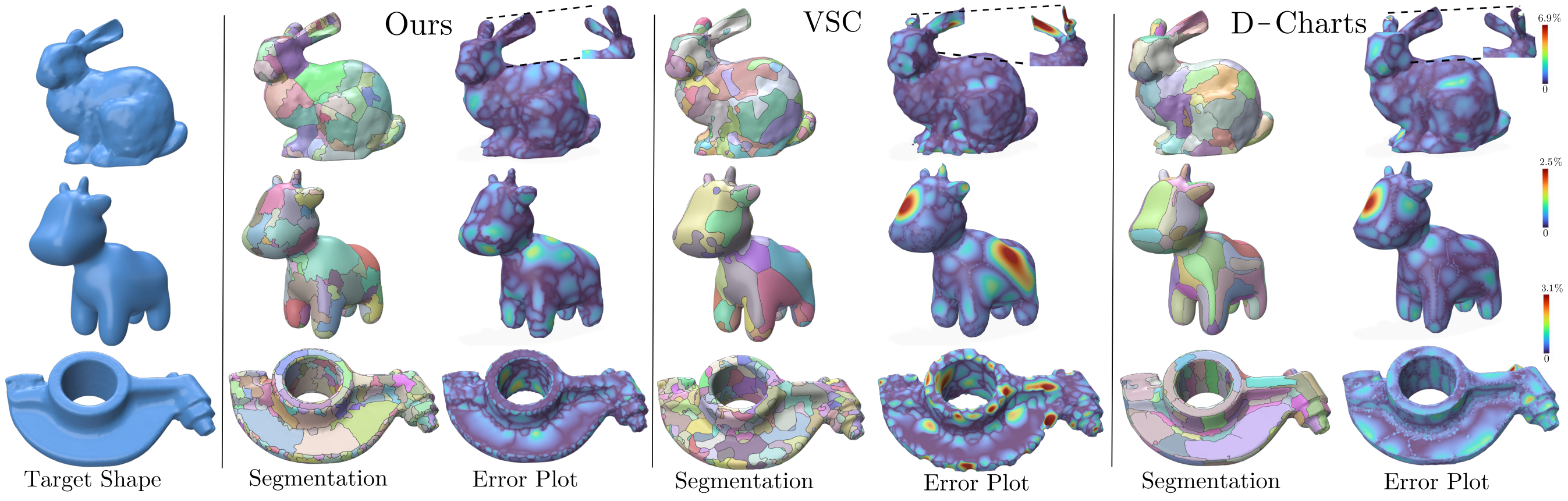}
\caption{Comparison of deviation plots for various meshes between our method, VSC and D-Charts. From Left to Right: Target Geometry, Segmentation computed by our method, corresponding simulated result with deviation from target visualized, Segmentation computed by VSC, corresponding Simulated result with deviation from target visualized, Segmentation computed by D-Charts, corresponding Simulated result with deviation from target visualized. The target mean curvatures are $H^* = 0.8$, $H^* = 1.8$ and $H^* = -0.02$ for the Bunny, Spot and Rocker meshes respectively.}
\label{fig:dev_results}
\end{figure*}
\section{Results and Discussion}

We present both fabricated as well as simulated results to show the efficacy of our piecewise-CMC segmentation method for fabricating freeform structures with fluidic shaping. We will also show how our method achieves better approximation to the input target geometry, compared to using standard developability or conformal distortion minimizing segmentation frameworks such as \textit{D-Charts} \cite{julius2005d} and \textit{Variational Surface Cutting (VSC)} \cite{sharp2018variational} respectively.

\subsection{Fabricated Results}

\begin{figure}[!t]
  \centering
  \includegraphics[width=\columnwidth]{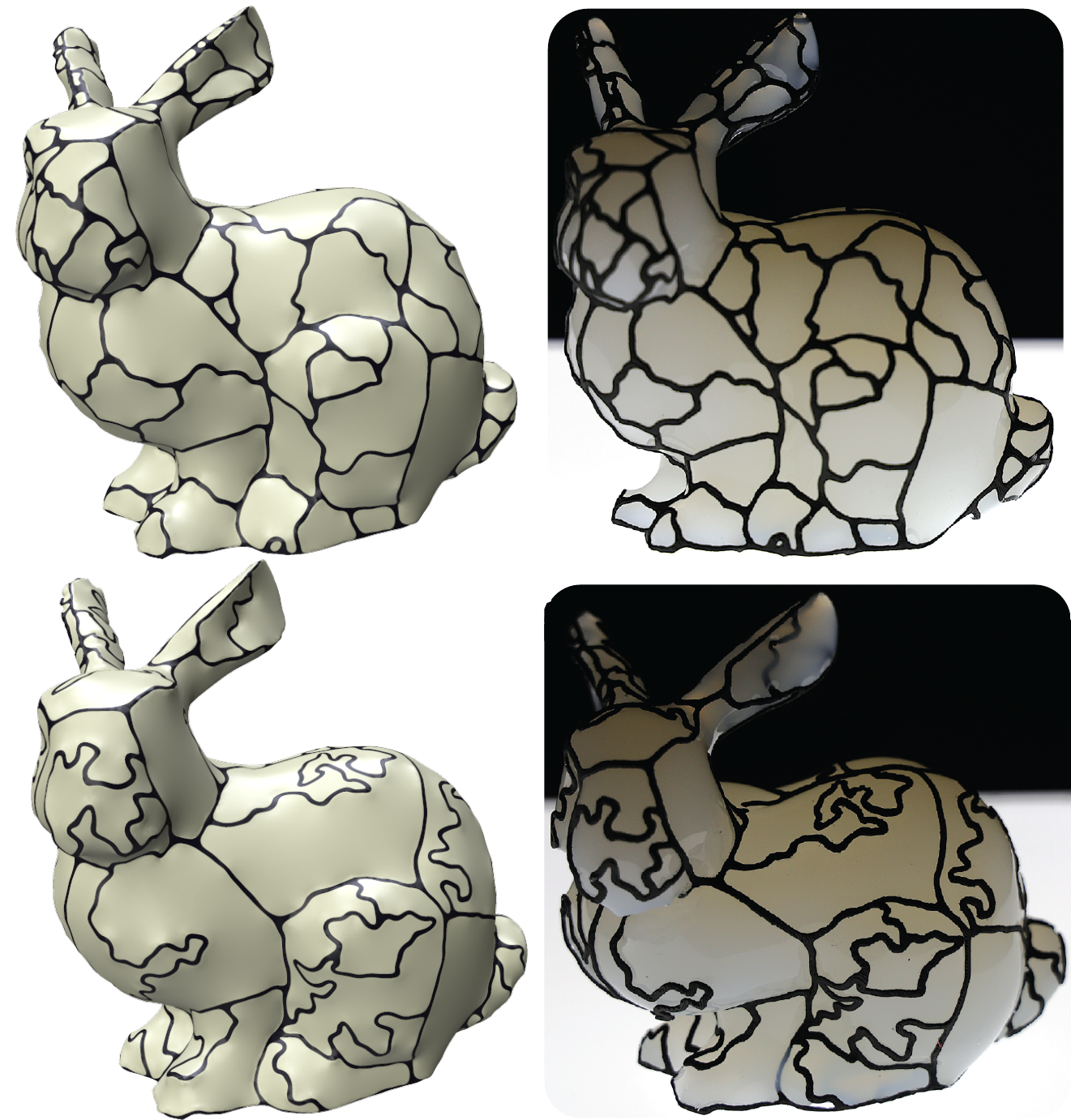}
  \caption{Comparison of compact VS non-compact segmentation and the corresponding fabricated outcomes. Left: Simulated Results. Right: Fabricated Results. Top row: Compact segmentation. Bottom row: Non-compact segmentation.}
  \label{fig:bunny_comp}
\end{figure}

We fabricated three Stanford bunny models and a prosthetic hand using the fluidic fabrication process. The blue bunny and red hand can be seen in the teaser image in Fig. \ref{fig:teaser}. The second and third bunnies are cream colored, shown in Fig. \ref{fig:bunny_comp}. The fabricated models show good agreement with the input target geometry. We evaluate them purely visually, and didn't take quantitative measurements by scanning the fabricated model to compute the differences. 

Fig. \ref{fig:bunny_comp} shows the cream colored bunnies fabricated using a compact and non-compact wireframe scaffold. On the left column, we showcase the simulation results, and the right column displays the corresponding fabricated models. We see good agreement between what the simulation predicts and the actual fabricated models. 

\subsection{Simulated Results}

We show some simulation only results for the piecewise-CMC segmentation method. After a segmentation is computed for the input target geometry, we run the P-CMC evolution described in Section \ref{subsec:cmc_evo}. This gives a good proxy for what the final fabricated surface would look like, allowing us to evaluate the efficacy of the segmentation without the need for physical fabrication. 
Fig. \ref{fig:sim_results} shows the simulated results for three different input target geometries - the Fandisk model, the Rocker model and Homer. We observe that the simulated surfaces closely follow the input target geometries, and mean curvature difference is concentrated on the seams. As an example, if we look at the Rocker model (middle row), we see the raw mean curvature is really positive on the sharp rims and negative in the saddle type regions. But most of the structure has mean curvature close to 0. The MC estimator of Section \ref{subsec:rep_mc} outputs an $H^* = -0.02$. 
Our segmentation hence places the seams in regions where mean curvature difference $H - H^*$ is high, and allows the rest of the structure to be smooth and have a mean curvature equal to -0.02. We can observe similar behavior in the Fandisk and Homer models as well.

\begin{figure}[!t]
    \centering
    \includegraphics[width=\columnwidth]{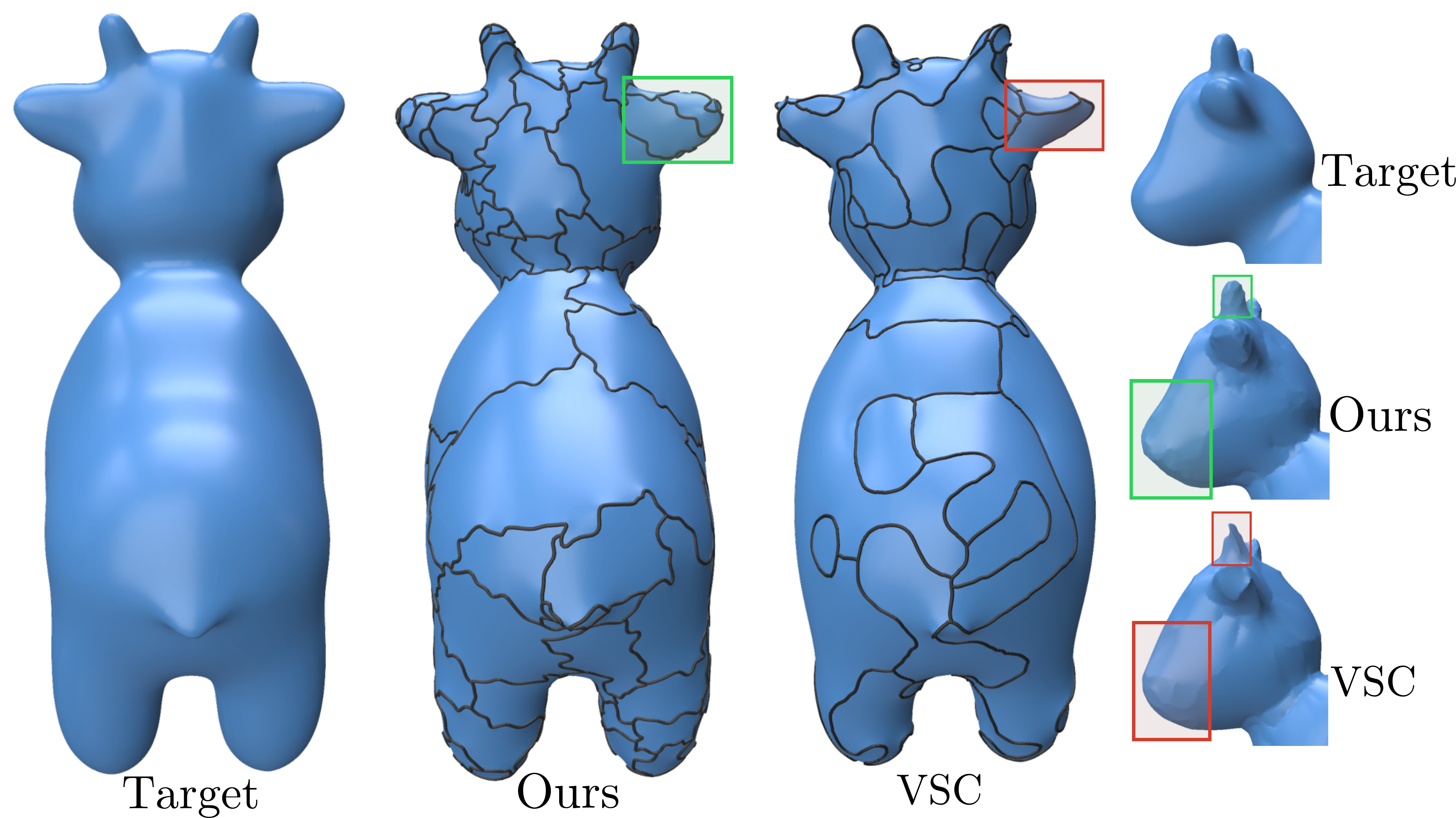}
    \caption{Comparison of our method with \textit{Variational Surface Cutting (VSC)}\cite{sharp2018variational}. Simulated results demonstrate our P-CMC segmentation outperforms VSC in preserving the target geometry.}
    \label{fig:vsc_comp}
\end{figure}

\subsubsection{Ground Truth Tests}

\begin{figure}[!t]
  \centering
  \includegraphics[width=\columnwidth]{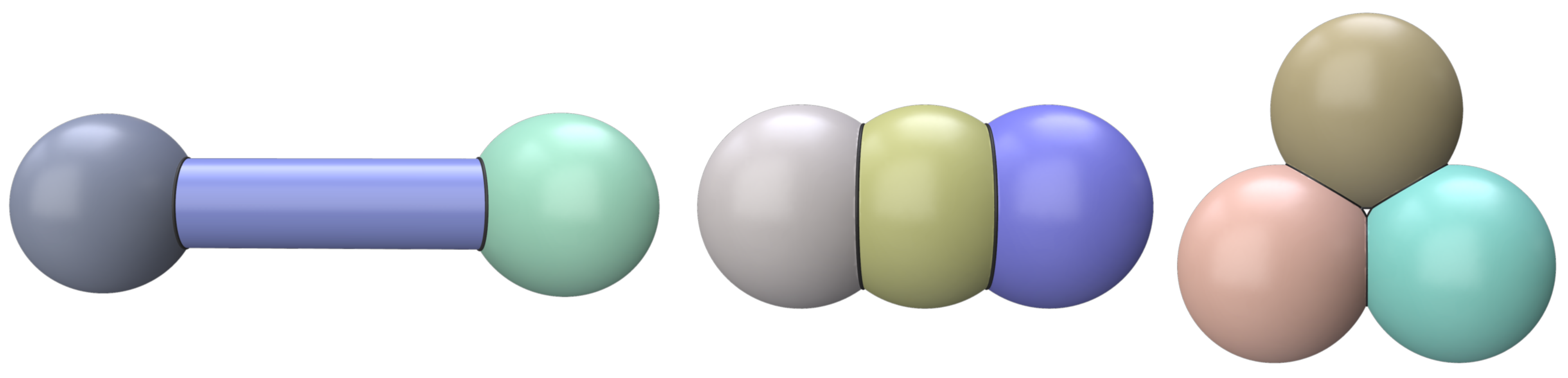}
  \caption{Segmentation of simple geometries into piecewise CMC surfaces. In this example, the target MC, $H^*=1.0$. The segment seams are shown in black. The evolution process of Section \ref{subsec:cmc_evo} has no problem converging to CMC surfaces that are saddle points of $E$, such as the major caps of the sphere seen in all examples.}
  \label{fig:simple_seg}
\end{figure}

We segment some geometries whose results are known to us beforehand -- hence they make for good ground truth tests. These models
are made up of the simplest CMC shapes such as spheres and cylinders.
We showcase 3 such models in Fig. \ref{fig:simple_seg}; the first one is a dumbbell where a cylinder connects two sphere caps, 
the second and third ones are 3 spheres fused together in different ways. Note that all spherical regions are major caps, and the CMC 
evolution procedure of Section~\ref{subsec:cmc_evo} converges to them without difficulty, which allows the segmentation to 
segregate them effectively.

\subsubsection{Comparisons with VSC}

We also show the efficacy of our piecewise-CMC segmentation method by comparing it with the VSC method. It is important to note that VSC optimizes for segments that minimize conformal scaling upon flattening to the plane. The goal is to demonstrate that the underlying physical process - \textit{Fluidic Shaping}; informs a certain type of geometry - \textit{CMC surfaces}; and it is paramount to account for it in the segmentation to achieve good target shape approximations. 

To make the comparison as fair as possible, we ensure that VSC produces a segmentation such that total seam length of the cuts is almost equal to that of our piecewise-CMC segmentation. We initialized VSC with the Normal Clustering scheme, as they did in the original paper \cite{sharp2018variational}. Fig. \ref{fig:vsc_comp} shows a qualitative comparison between the two methods. 
We perform the P-CMC segmentation with a target $H^*$ computed using the method described in Section \ref{subsec:rep_mc}. Once we obtain both segmentations, we run the P-CMC evolution on the input target geometry with seams from each segmentation as the fixed boundary conditions and the target $H^*$. We can see the resulting simulated surfaces in the middle and right in Fig. \ref{fig:vsc_comp}.
The VSC segmentation results in a simulated surface that deviates more from the target geometry, as can be seen in the ear and head regions. The sides of Spot are also more inflated than they should be. For a better visualization, we urge the reader to refer to Fig. \ref{fig:dev_results}, middle row.

\begin{table}[!t]
  \centering
  \caption{Deviation of the simulated piecewise-CMC surfaces from the target meshes of Fig.~\ref{fig:dev_results}, as a percentage of the bounding box diagonal of the target mesh. P95 is the 95th percentile. Lower is better.}
  \label{tab:dev_results}
  \setlength{\aboverulesep}{0pt}
  \setlength{\belowrulesep}{0pt}
  \renewcommand{\arraystretch}{1.3}
  \begin{tabularx}{\columnwidth}{l|l*{4}{>{\centering\arraybackslash}X}}
    \toprule
    Mesh & Method & Mean & RMS & P95 & Max \\
    \midrule
    \rowcolor{oursrow}
    \cellcolor{white}\multirow{3}{*}{Bunny}  & Ours & 0.20 & 0.31 & 0.71 & 1.34 \\
                                             & VSC  & 0.35 & 0.87 & 1.61 & 6.97 \\
                                             & D-Charts & 0.23 & 0.33 & 0.69 & 3.5 \\
    \midrule
    \rowcolor{oursrow}
    \cellcolor{white}\multirow{3}{*}{Spot}   & Ours & 0.15 & 0.22 & 0.49 & 0.95 \\
                                             & VSC  & 0.28 & 0.47 & 1.22 & 2.51 \\
                                             & D-Charts & 0.17 & 0.25 & 0.49 & 1.64 \\
    \midrule
    \rowcolor{oursrow}
    \cellcolor{white}\multirow{3}{*}{Rocker} & Ours & 0.11 & 0.16 & 0.35 & 0.78 \\
                                             & VSC  & 0.22 & 0.40 & 0.86 & 3.13 \\
                                             & D-Charts & 0.14 & 0.18 & 0.33 & 2.12 \\
    \bottomrule
  \end{tabularx}
\end{table}

Fig. \ref{fig:dev_results} also showcases the Bunny and Rocker model comparisons in the top and bottom rows. In each of the two models, our method produces a better overall global target shape approximation compared to VSC. Table \ref{tab:dev_results} provides the quantitative deviations of the simulated surfaces from the target meshes. Across all metrics, we observe that our method consistently achieves lower deviations than VSC. We can hence claim that a segmentation tailormade for the underlying physical process of fluidic shaping leads to superior target shape approximations, compared to a general purpose segmentation.

There are some regions on these target geometries where VSC can perform better, because it creates more cuts in those regions (see Fig. \ref{fig:dev_results}, top row). In the Bunny model for instance, the VSC segmentation places more cuts in the back of the bunny, which helps to reduce the deviation in that area. However, we can interpret this as VSC spending more seam budget on those regions at the expense of other areas, such as the ears, which are heavily distorted.
Our approach on the other hand, is guaranteed to keep the RMSE error between the CMC surface and the target patch below a user defined threshold. This ensures no particular region is favored, and hence ensures a better global shape approximation. This is also important for certain applications where we want as much smoothness as possible and minimum amount of 3D printed frame necessary to maintain a good approximation to the target. 

\subsubsection{Comparisons with D-Charts}

We also compare our piecewise-CMC segmentation method with D-Charts. D-Charts is a segmentation method that aims to partition the surface into regions that are as developable as possible. It requires $k$ as input, where $k$ is the number of desired segments. We
pick $k$ such that the total seam is length is close to that of our piecewise-CMC segmentation for a fair comparison. As we saw with the VSC method, the D-Charts segmentation also results in simulated surfaces that deviate more from the target geometry, particularly in regions where the segmentation places fewer cuts (Fig. \ref{fig:dev_results}). 

\section{Limitations and Future Work}

\begin{figure}[!t]
    \centering
    \includegraphics[width=\columnwidth]{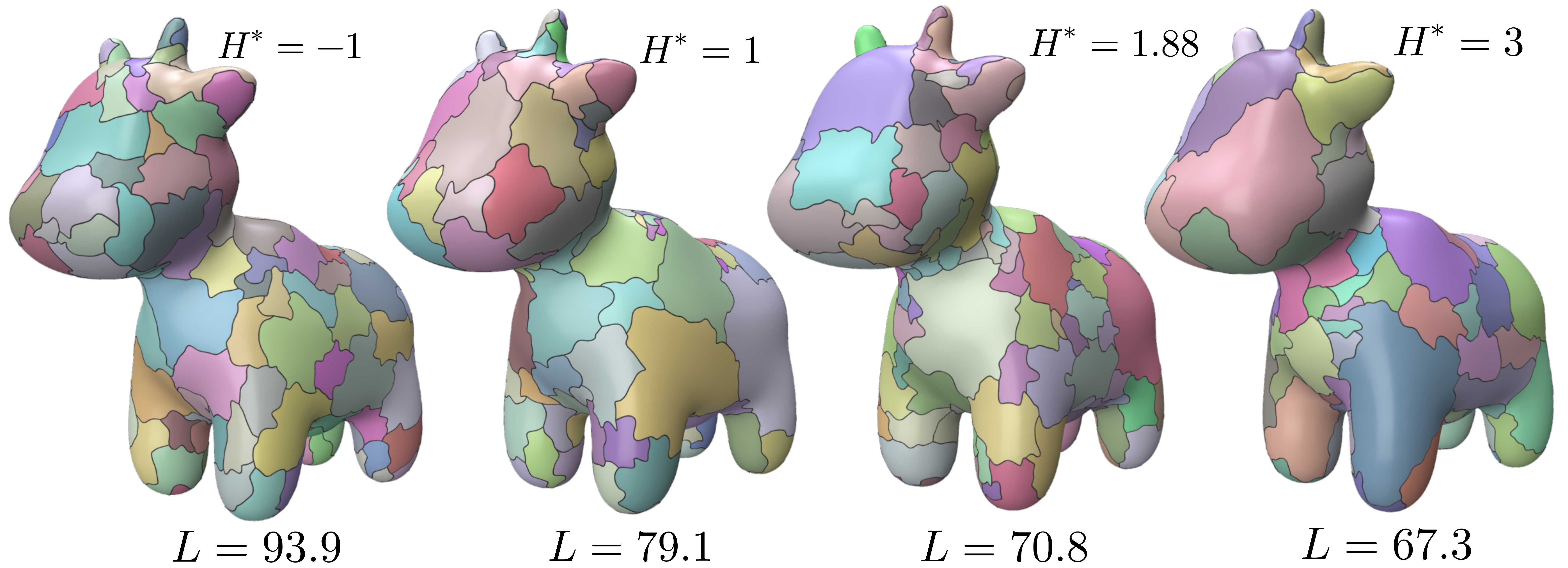}
    \caption{Segmenting the mesh with different target mean curvatures $H^*$. $L$ denotes the total seam length produced by the segmentation. $H^* = 1.88$ for the third result from left was computed using the MC estimator of Section \ref{subsec:rep_mc}.}
    \label{fig:H_comp}
\end{figure}

Fig. \ref{fig:H_comp} illustrates the effect of different target mean curvatures on the segmentation results. The total seam length $L$ varies with the choice of $H^*$, highlighting the sensitivity of the segmentation to the target curvature. Our MC estimator heuristic of Section \ref{subsec:rep_mc} provides a practical way to estimate $H^*$ from the target geometry, but it may not always yield the optimal value for minimizing the total seam length, as can be seen in the third result from the left, where the estimated $H^* = 1.88$ does not achieve the minimum $L$.

The greedy merging combined with the compactness criterion prevents certain merges that reduce compactness in the current merge cycle, but another merge later on might improve the overall compactness. This can lead to segmentation outcomes that need more patches than necessary. A potential solution would be use hypergraphs that can encode 3 or 4 way merges and have a certain \textit{look-ahead} capability to evaluate the impact of future merges on the overall compactness.

The underlying physics of the fabrication process allows for only one global pressure or mean curvature $H^*$, as we explored in Section \ref{subsec:phys_conn}. Since most target geometries have varying mean curvatures, this leads to the creation of more scaffolding in regions where the local mean curvature deviates from the global value $H^*$. An interesting future work would be to allow multiple mean curvatures, by separating the fluid bodies through internal routing in the structure. 

Building on the work from \citet{bhundiya2023bend} and \citet{Faruqi2025WireBend}, future research could also explore the integration of 3D bend-forming techniques with our piecewise-CMC segmentation method to fabricate the 3D wireframe scaffolds much faster. These metal scaffolds can also serve as durable reinforcements and make the cured polymer structures stronger and more resilient.

\section{Conclusion}

We presented a piecewise CMC segmentation method that extends fluidic shaping from single optical surfaces to freeform 3D geometry. Given a target mean curvature, our method grows segments through graph contractions until a patch can no longer be approximated by a CMC surface within a user-defined tolerance, concentrating the non-CMC features of the target onto a sparse set of seams. These seams become a 3D printed frame that pins the polymer interface, while surface tension shapes the remaining bulk into smooth patches without layer-wise deposition or post-processing. Fabricated prototypes visually agree with simulation, and our simulated surfaces approximate the targets with consistently lower deviation than a general-purpose segmentation of comparable seam length.

\bibliographystyle{ACM-Reference-Format}
\bibliography{references}

\appendix

\section{Critical Points of $E$ are CMC Surfaces}
\label{app:crit_cmc}

Using standard discrete differential geometry \cite{meyer2003discrete, crane2018discrete}, the gradients of the two terms in Equation~\eqref{eq:energy_H} are
\begin{equation}
    \nabla A(\bm{X}) = 2\bm{M}\bm{H}\bm{n}, \qquad
    \nabla V(\bm{X}) = \bm{M}\bm{n},
    \label{eq:ddg_grads}
\end{equation}
where $\bm{M}$ is the lumped mass matrix of vertex areas, $\bm{H}$ the vertex mean curvatures and $\bm{n}$ the vertex normals, all of which depend on the current geometry $\bm{X}$. Differentiating Equation~\eqref{eq:energy_H} and substituting Equation~\eqref{eq:ddg_grads} gives a clean expression for the energy gradient,
\begin{equation}
    \nabla E(\bm{X}) = 2\bm{M}\left(\bm{H} - \bm{H}^*\right)\bm{n}.
    \label{eq:grad_E}
\end{equation}
Since $\bm{M}$ is diagonal with strictly positive entries and the $\bm{n}_i$ are unit vectors, $\nabla E(\bm{X}) = 0$ holds if and only if $H_i = H^*$ at every interior vertex. Critical points of $E$ are thus precisely the discrete CMC surfaces of target curvature $H^*$. Boundary vertices are pinned by the wireframe scaffold and are held fixed throughout, so the gradient is only ever enforced on the interior degrees of freedom.

\section{Physical Connection: Surface Tension, Pressure and CMC Equilibrium}
\label{app:phys_conn}

In this section, we derive the static force balance of Equation~\eqref{eq:equilibrium_area_grad}, and then shows that the critical points of the geometric energy $E$ of Equation~\eqref{eq:energy_H} satisfy it.

\subsection{Surface tension as an area-minimizing force}

A surface with surface tension coefficient $\alpha$ stores an interfacial
energy proportional to its area around a point $x$ \cite{lautrup2010surface_tension},
\begin{equation}
    E_{\mathrm{surf}}(x) = \alpha A(x),
\end{equation}
so any deformation that increases area increases energy. The surface-tension
force per unit area is proportional to the negative gradient of area with
respect to the surface position $x$,
\begin{equation}
    f_{\mathrm{st}}(x) \;\bm{\propto}\; -\alpha\,\frac{\partial A(x)}{\partial x}.
\end{equation}
For smooth surfaces, the first variation of area is the mean-curvature
normal,
\begin{equation}
    \frac{\partial A(x)}{\partial x} = 2H(x)\,\bm{n}(x),
    \label{eq:area_grad_continuum}
\end{equation}
so surface tension acts along $-2\alpha H\bm{n}$, i.e.\ opposite the outward normal, moving the surface in the direction that reduces area the fastest.

\subsection{Pressure as a pushback against contraction}

Left unopposed, this area-minimizing force would collapse any closed surface.
The incompressible fluid trapped inside resists, pushing back against the
contraction with a force
\begin{equation}
    f_{\mathrm{p}}(x) \;\bm{\propto}\; \Delta p \, \bm{n}(x),
\end{equation}
where $\Delta p$ is the pressure excess at $x$.\footnote{Without this term a
sphere would collapse to a point and every fixed-boundary surface would relax to a minimal surface. This is consistent with the Young--Laplace law: $\Delta p
= 0 \implies H = 0$. Physically, a minimal surface is open and encloses no
fluid, so the pressure on either side is equal and there is no pressure discontinuity to sustain a nonzero mean curvature.} 

\subsection{Static equilibrium of the interface}

A surface-only dynamics model \cite{ishida2017hgf} combining the above two effects is
\begin{equation}
    \rho_s \frac{d^2 x}{dt^2}
    = -\alpha\,\frac{\partial A(x)}{\partial x} + \Delta p \, \bm{n},
    \label{eq:phys_dyn}
\end{equation}
where $\rho_s$ is an effective surface mass density. At static equilibrium the acceleration vanishes, and Equation~\eqref{eq:phys_dyn} reduces to a balance between the inward pull of surface tension and the outward push of pressure, which is exactly Equation~\eqref{eq:equilibrium_area_grad} of the main text.

\makeatletter\@secpenalty=\z@\makeatother
\subsection{Recovering the force balance from the geometric model}
\makeatletter\@secpenalty=-300\makeatother

It remains to show that the critical points of $E$ satisfy Equation~\eqref{eq:equilibrium_area_grad}. Expanding the stationarity condition $\nabla E(\bm{X}) = 0$ for the energy in Equation~\eqref{eq:energy_H} gives
\begin{equation}
    \nabla A(\bm{X}) - 2H^* \nabla V(\bm{X}) = 0
    \quad\Longrightarrow\quad
    \nabla A(\bm{X}) = 2H^* \nabla V(\bm{X}).
    \label{eq:crit_point}
\end{equation}
Multiplying both sides by the surface tension coefficient $\alpha$ and substituting the discrete gradients of Equation~\eqref{eq:ddg_grads},
\begin{equation}
    \alpha \left(2\bm{M}\bm{H}\bm{n}\right) = \left(2\alpha H^*\right) \bm{M}\bm{n}.
    \label{eq:alpha_scaled}
\end{equation}
The Young--Laplace law of Equation~\eqref{eq:yl} identifies the bracketed constant on the right hand side as the pressure jump $\Delta p$ sustained by an interface of mean curvature $H^*$. Substituting it into Equation~\eqref{eq:alpha_scaled} and cancelling the mass matrix $\bm{M}$, which is diagonal with strictly positive entries, leaves
\begin{equation}
    \alpha \left(2\bm{H}\bm{n}\right) = \Delta p \, \bm{n},
\end{equation}
which, by the continuum area gradient of Equation~\eqref{eq:area_grad_continuum}, is exactly the force balance of Equation~\eqref{eq:equilibrium_area_grad}, with the target curvature $H^*$ and the pressure jump $\Delta p$ related through Equation~\eqref{eq:yl}.

\section{Constrained Quadric Fitting}
\label{app:quadric}

This section derives the constrained least squares system used to fit quadrics to small segments in Section~\ref{subsec:quadrics}.

\subsection{Local height-graph coordinates}

Given a vertex \(\bm{p}\) representing the centroid of a segment, we collect a neighborhood of nearby sample points \(\{\bm{P}_k\}\) and approximate the local surface by a quadratic height field in the frame \((\bm{t}_1,\bm{t}_2,\bm{n})\) at \(\bm{p}\). Each neighboring point \(\bm{P}_k\) is expressed in local coordinates \((u_k,v_k,h_k)\) via
\begin{equation}
\begin{bmatrix}
u_k\\
v_k\\
h_k
\end{bmatrix}
=
\begin{bmatrix}
\text{---}\, \bm{t}_1^\top \,\text{---}\\
\text{---}\, \bm{t}_2^\top \,\text{---}\\
\text{---}\, \bm{n}^\top \,\text{---}
\end{bmatrix}
(\bm{P}_k - \bm{p}),
\end{equation}
which gives a height-graph representation of the local geometry around \(\bm{p}\). The patch is modeled by the quadratic of Equation~\eqref{eq:quadric_model}, with unknown coefficients
\(
\bm{\theta} = [a,b,c,d,e,f]^\top.
\)

\subsection{Apex constraints}

We want the central vertex \(\bm{p}\) to correspond to a critical point of the fitted graph, and impose zero first derivatives at the local origin to accomplish this:
\begin{equation}
\frac{\partial w}{\partial u}(0,0)=0,
\qquad
\frac{\partial w}{\partial v}(0,0)=0.
\end{equation}
For the quadratic model,
\begin{equation}
\frac{\partial w}{\partial u} = 2au + bv + d,
\qquad
\frac{\partial w}{\partial v} = bu + 2cv + e,
\end{equation}
so the apex constraints imply \(d = 0\) and \(e = 0\).

\subsection{Prescribing the mean curvature}

With the apex constraints in place, the mean curvature at the apex takes a simple form. For a height graph \(z = w(u,v)\), the mean curvature at a critical point is
\begin{equation}
H = \frac{w_{uu} + w_{vv}}{2}.
\end{equation}
For the quadratic model the second partials are \(w_{uu} = \frac{\partial^2 w}{\partial u^2} = 2a\) and \(w_{vv} = \frac{\partial^2 w}{\partial v^2} = 2c\), so
\begin{equation}
H = a + c,
\end{equation}
and prescribing a target mean curvature \(H^\ast\) yields the linear constraint \(a + c = H^\ast\). We enforce it by eliminating a degree of freedom, introducing \(\delta = a - c\) so that
\begin{equation}
a = \frac{H^\ast + \delta}{2},
\qquad
c = \frac{H^\ast - \delta}{2}.
\end{equation}

\subsection{The reduced least squares system}

Substituting these expressions, together with the apex constraints \(d = e = 0\), into the quadratic model gives
\begin{equation}
w(u,v)
=
\frac{H^\ast}{2}(u^2+v^2)
+
\frac{\delta}{2}(u^2-v^2)
+
buv + f.
\end{equation}
Treating \(H^\ast\) as fixed, the unknowns reduce to \(\bm{\theta} = [\delta,b,f]^\top\). Each sample point \((u_k,v_k,h_k)\) then contributes the linear relation
\begin{equation}
\frac{\delta}{2}(u_k^2-v_k^2) + b u_k v_k + f
\approx
h_k - \frac{H^\ast}{2}(u_k^2+v_k^2),
\end{equation}
and stacking all samples yields a linear least-squares system \(\bm{B}\bm{\theta} \approx \bm{y}\), which we solve for the constrained quadric coefficients. This produces a local quadratic patch that both fits the data and satisfies the prescribed apex and mean-curvature constraints.

\section{Minimizers of the Robust CMC Energy are Stable}
\label{app:stability}

The outer search of Section~\ref{subsec:cmc_evo} reaches Newton's basin by repeatedly minimizing the energy $F(\bm{X};t)$ (to compute the residual $H^* - H$) of Equation~\eqref{eq:pan_energy}, from \citet{pan2012robust}. We show in the following that those inner minimizations produce physically \textit{stable} CMC surfaces: among all nearby surfaces holding the same boundary and enclosing the same volume, it is the one of least area. This property decides whether a CMC patch can be fabricated at all, since it is the shape the polymer settles into once the frame is filled.

The physical reading of this result is as follows. A fixed volume of polymer pinned by the wireframe settles into the shape that minimizes interfacial area at that volume. Such a shape is at equilibrium and no small volume-preserving disturbance can lower its energy, so it doesn't carry any unstable modes or directions.

\begin{definition}[Stability]
\label{def:stability}
A CMC surface $\Sigma_0$ with pinned boundary is \emph{stable} if $A(\Sigma) \geq A(\Sigma_0)$ for every nearby surface $\Sigma$ that shares the boundary of $\Sigma_0$ and encloses the same volume, $V(\Sigma) = V(\Sigma_0)$.
\end{definition}

The energy of Equation~\eqref{eq:pan_energy} pairs an extended centroidal Voronoi tessellation term with the volume functional $F_{\mathrm{vol}} = 3V$, where $V$ is the signed enclosed volume. For a fixed sampling, the ECVT term is asymptotically proportional to the squared area \cite{pan2012robust}, which gives the smooth surrogate
\begin{equation}
    F(\Sigma; t) = c\,A(\Sigma)^2 + 3t\,V(\Sigma),
    \qquad c > 0,
    \label{eq:smooth_cvt}
\end{equation}

\subsection{Minimizing the energy minimizes area at fixed volume}
\label{subsec:area_min}

\begin{proposition}[Stability of the computed surfaces]
\label{prop:stable}
Let $\Sigma_0$ be a local minimizer of $F(\cdot\,; t)$ of Equation~\eqref{eq:smooth_cvt} over surfaces with the boundary pinned, with $A(\Sigma_0) > 0$. Then $\Sigma_0$ is stable in the sense of Definition~\ref{def:stability}.
\end{proposition}

\begin{proof}
Let $\Sigma$ be any nearby surface that shares the boundary of $\Sigma_0$ and encloses the same volume, $V(\Sigma) = V(\Sigma_0)$. Since $\Sigma_0$ is a local minimizer of $F$,
\begin{align}
    0 \ &\leq \ F(\Sigma) - F(\Sigma_0) \nonumber \\
      &= \ c\left[A(\Sigma)^2 - A(\Sigma_0)^2\right]
         + 3t\left[V(\Sigma) - V(\Sigma_0)\right] \nonumber \\
      &= \ c\left[A(\Sigma)^2 - A(\Sigma_0)^2\right],
    \label{eq:energy_gap}
\end{align}
where the volume terms cancel identically because the perturbation preserves volume. Since $c > 0$, this gives $A(\Sigma)^2 \geq A(\Sigma_0)^2$, and as both areas are positive, $A(\Sigma) \geq A(\Sigma_0)$. If $\Sigma_0$ is a strict local minimizer the inequalities are strict, making it the unique area minimizer among the nearby surfaces at its own volume.
\end{proof}

\subsection{Saddle points of $E$ can still be stable}
\label{subsec:saddles_stable}

Section~\ref{subsec:cmc_energy} describes the major spherical cap as a saddle point of the physically based energy $E = A - 2H^*V$, while Proposition~\ref{prop:stable} asserts that everything the root finder returns is stable. We will establish a connection between the two energies, $E$ and $F$ in the following.

A saddle point of $E$ is unconstrained: it permits deformations that change the enclosed volume. For a spherical cap on a fixed boundary circle, the unstable direction is the one that slides the surface through the family of caps on that circle, growing the minor cap into the major one. That motion changes the enclosed volume, so it is inadmissible the moment the fluid body is fixed. Removing it drops the index -- the number of independent directions along which the surface can shed energy, from one to zero.

On the set where $V(\Sigma) = V_0$, the energy can be written as
\begin{equation}
    E(\Sigma) = A(\Sigma) - 2H^* V_0
    \label{eq:E_fixed_vol}
\end{equation}
so minimizing area at fixed volume and minimizing $E$ become the same problem. A surface can accordingly be a saddle of $E$ in the unconstrained problem and a minimizer of that very same $E$ once the volume is pinned.

This is what makes the major caps of Fig. \ref{fig:simple_seg} fabricable rather than merely computable. Each is produced as a local minimizer of $F$ and is therefore stable by Proposition~\ref{prop:stable}: a genuine equilibrium of the filled frame, not a configuration the fluid would immediately leave.

\subsection{Numerical verification}
\label{subsec:stability_numerics}

\begin{figure}[!t]
  \centering
  \includegraphics[width=\columnwidth]{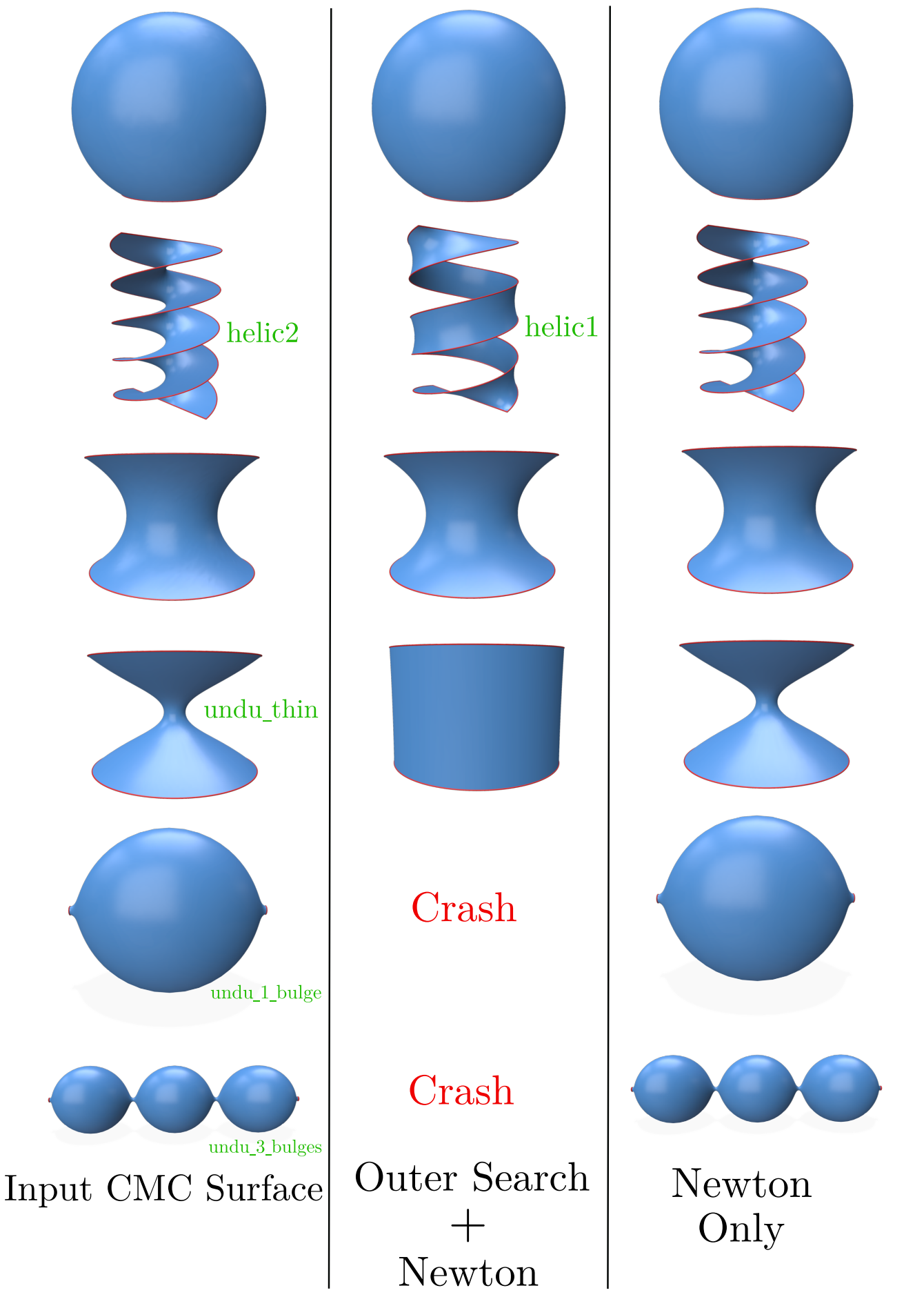}
  \caption{Left: input CMC surfaces, with their fixed boundaries in red. Center: the result of the outer search followed by Newton refinement (Section~\ref{subsec:cmc_evo}). Right: the result of Newton refinement alone. Newton alone stays on the input whether or not it is stable, while the root finder returns only stable surfaces (Table~\ref{tab:stability}). }
  \label{fig:stability}
\end{figure}

\begin{table}[!t]
  \centering
  \caption{Number of negative fixed-volume modes of the surfaces in Fig.~\ref{fig:stability}. A surface with at least one such mode can lower its area at fixed volume and is therefore unstable.}
  \label{tab:stability}
  \begin{tabular}{lc}
    \toprule
    Mesh & Number of negative fixed-volume modes \\
    \midrule
    Major cap                  & 0 \\
    \texttt{helic1}            & 0 \\
    \texttt{helic2}            & 1 \\
    Catenoid                   & 0 \\
    \texttt{undu\_thin}        & 1 \\
    Cylinder                   & 0 \\
    \texttt{undu\_1\_bulge}    & 0 \\
    \texttt{undu\_3\_bulges}   & 4 \\
    \bottomrule
  \end{tabular}
\end{table}

Proposition~\ref{prop:stable} can be checked empirically on the meshes.
Fig.~\ref{fig:stability} shows runs of our evolution process (Section \ref{subsec:cmc_evo}) on a set of known CMC surfaces, 
once with the full two-phase solver and once with Newton's method alone. Table~\ref{tab:stability} reports the negative fixed-volume 
modes for every surface. These modes can be computed via Surface Evolver, by setting a fixed volume constraint and evaluating the Hessian of energy $E$. 

Every surface the root finder (outer search + Newton) returns has 0 negative fixed-volume modes, as Proposition~\ref{prop:stable} predicts. Newton's method alone is indifferent to 
stability: it stays on the CMC surfaces \texttt{helic2}, \texttt{undu\_thin} and \texttt{undu\_3\_bulges}, which carry one, one and four negative modes 
respectively. A frame filled with polymer fluid wouldn't be able to hold these. On \texttt{helic2} and \texttt{undu\_thin}, the root finder instead 
settles on a stable CMC surface with the same boundary, \texttt{helic1} and a cylinder respectively. It fails on both bulged unduloids, including 
the stable \texttt{undu\_1\_bulge}, due to an issue in computing a stable $t_0$ (Equation \ref{eq:t0}) to start the inner solve for the root finder (Algorithm \ref{alg:two_path}). 
The empirical evidence thus shows that when the root finder does converge, it returns a stable CMC surface as per Definition~\ref{def:stability}.

\begin{figure}[!t]
  \centering
  \includegraphics[width=\columnwidth]{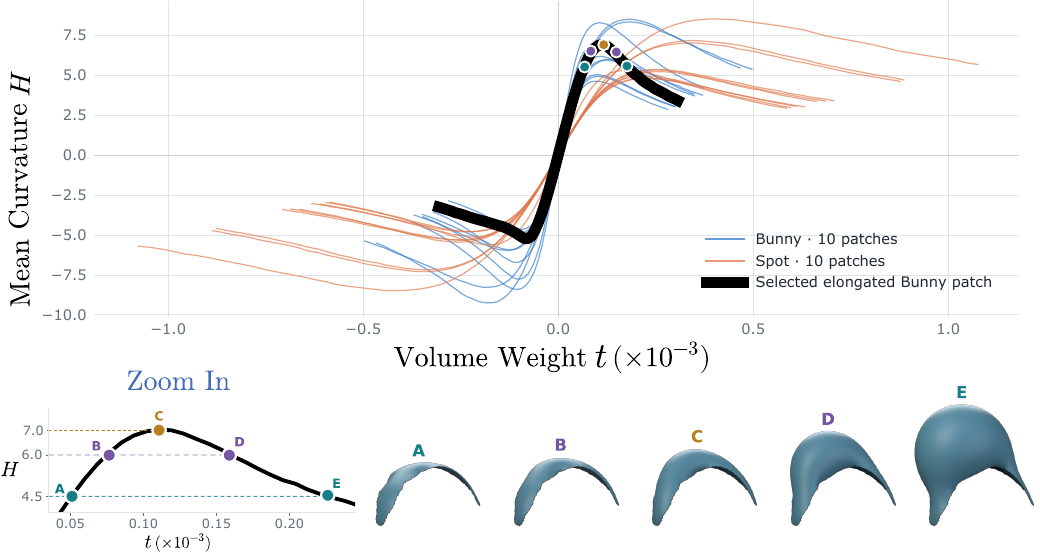}
  \caption{Top: Mean curvature $H(t)$ of the minimizer of $F(\cdot\,;t)$ as a function of the volume weight $t$ of Equation~\eqref{eq:pan_energy}, for randomly sampled segments on the Bunny (blue) and Spot (orange) models. Curve in black shows selected patch. Bottom: Zoomed in version of the curve and some snapshots of the corresponding CMC surfaces with increasing $t$. We can see "small" and "large" solutions akin to minor and major caps of a sphere respectively. }
  \label{fig:tH}
\end{figure}

Since every evaluation of the residual $g(t)$ of Equation~\eqref{eq:residual_t} requires a minimization of $F$, the root finder walks along the 
curve $H(t)$ of stable surfaces. Fig.~\ref{fig:tH} plots this curve for segments of the Bunny and Spot models. The curve is monotone through $t = 0$, 
but turns over at an extremum, beyond which $|H|$ decreases as $|t|$ grows. A target $H^*$ short of the extremum is therefore 
reached at two values of $t$, one on either side of it, and a target beyond it is not reached at all. Near the extremum the slope of $H(t)$ vanishes, 
so a secant step there can be thrown far along $t$ and onto the other branch. The search of Appendix~\ref{app:root_search} guards against both 
effects: it bounds each secant step by a trust radius, and it runs a separate path on each side of its starting weight, keeping the nearer root.

\section{Two-Path Root Search}
\label{app:root_search}

\begin{algorithm}[!t]
  \caption{Two-path search for $t, \mathrm{s.t.}\ g(t) = H(t) - H^* = 0$}
  \label{alg:two_path}
  \begin{algorithmic}[1]
    \Require surface $\bm{X}$ with pinned boundary, target $H^*$, tolerance $\tau$, probe step $\epsilon$, trust radius $\delta$
    \State $t_0 \gets$ starting weight of $\bm{X}$, Equation~\eqref{eq:t0}
    \State \textbf{if} $|g(t_0)| \le \tau$ \textbf{then return} $t_0$
    \State $\mathit{dirs} \gets \{-1, +1\}$ \Comment{deflate, inflate}
    \State \textbf{if} $t_0 = 0$ \textbf{then} $\mathit{dirs} \gets \{\operatorname{sign} H^*\}$ \Comment{minimal input}
    \For{$d \in \mathit{dirs}$} \Comment{one path per direction}
      \State $t_a \gets t_0$, \ $t_b \gets t_0 + d\,\epsilon$
      \Repeat
        \If{a bracket on the root is known}
          \State $t \gets$ Illinois step on the bracket
        \Else
          \State $t \gets$ secant step from $t_a, t_b$, with $|t - t_b| \le \delta$
        \EndIf
        \State keep $t$ on its own side: $d\,(t - t_0) \ge 0$
        \State evaluate $g(t)$, warm-started from the surface at $t_b$
        \State $t_a \gets t_b$, \ $t_b \gets t$
      \Until{$|g(t_b)| \le \tau$ or the evaluation budget is spent}
      \State $t_d \gets t_b$
    \EndFor
    \State \Return the converged $t_d$ closest to $t_0$
  \end{algorithmic}
\end{algorithm}

Algorithm~\ref{alg:two_path} demonstrates the outer search of Section~\ref{subsec:cmc_evo}. Initially, it computes a volume weight $t_0$ that gives the closest CMC surface to the input geometry, as in \cite{pan2012robust}
\begin{equation}
    t_0 = -\frac{1}{n_I} \sum_{i} \frac{\langle \nabla_i F_{\mathrm{area}},\, \nabla_i F_{\mathrm{vol}} \rangle}{\|\nabla_i F_{\mathrm{vol}}\|^2},
    \label{eq:t0}
\end{equation}
where $n_I$ is the number of interior vertices, $F_{\mathrm{area}}$ and $F_{\mathrm{vol}}$ are the area and volume functionals, and $\nabla_i$ denotes the gradient with respect to the $i$-th vertex.

Since $H(t)$ is not monotone (Fig.~\ref{fig:tH}), the target can have a root on either side of $t_0$, so we search both sides with two independent paths: one deflating the surface ($d = -1$, decreasing $t$) 
and one inflating it ($d = +1$). For an input which is a minimal surface or close to one, $t_0 = 0$ and already sits at $H = 0$, so only the path toward the sign of $H^*$ is run. Each path starts from a probe a small step $\epsilon$ away from $t_0$, 
and runs the secant or Illinois steps of Section~\ref{subsec:cmc_evo}. The trust radius $\delta$ caps how far a single secant step may move $t$; it grows after a step that 
reduces $|g|$ and halves after one that does not. A path that finds no root on its side simply fails, and the other path covers it. Of the paths that converge, we keep the one whose computed $t$ is closest to $t_0$. This
gives us a surface whose average mean curvature is $\approx$ $H^*$ that is closest to the input geometry.
If a solution within tolerance $\tau$ isn't reached after a certain number of iterations, the root finder terminates and reports failure.

\end{document}